\documentclass[journal=tches,final,nohyperref]{iacrtrans}

\usepackage{graphics}
\usepackage{epsfig} 
\usepackage{amssymb}
\usepackage{amsmath}

\usepackage{dsfont}
\usepackage{t1enc}
\usepackage{multirow}
\usepackage{tabularx}
\usepackage{fancybox}
\usepackage{epic}
\usepackage{color}
\usepackage{graphicx}
\usepackage{tikz}
\usetikzlibrary{positioning,chains,fit,shapes,calc}
\usepackage{pgfplots}
\usepackage{booktabs}
\usepackage{array}
\usepackage{url}
\usepackage{include/qian}
\usepackage{listings}
\usepackage[strings]{underscore}
\usepackage{diagbox}
\usepackage{booktabs}
\usepackage{caption}
\usepackage{subcaption}
\usepackage{standalone}
\usepackage{threeparttable}
\usepackage{numprint}
\usepackage{siunitx}
\usepackage[
n, 
advantage,  
operators,  
sets, 
adversary,  
landau,    
probability, 
notions,      
logic,     
ff,     
mm,  
primitives,  
events,
complexity,
oracles,   
asymptotics, 
keys 
]{cryptocode}

\usepackage[most]{tcolorbox}

\pgfplotsset{compat=1.18}

\usepackage{xcolor}   
 
\definecolor{codegreen}{rgb}{0,0.6,0}
\definecolor{codegray}{rgb}{0.5,0.5,0.5}
\definecolor{codepurple}{rgb}{0.58,0,0.82}
\definecolor{backcolour}{rgb}{0.95,0.95,0.92} 

\definecolor{myred}{RGB}{255,0,0}

\newcommand{\myred}[1]{#1}

\lstdefinestyle{mystyle}{
    backgroundcolor=\color{backcolour},   
    commentstyle=\color{codegreen},
    keywordstyle=\color{magenta},
    numberstyle=\tiny\color{codegray}, 
    stringstyle=\color{codepurple},
    basicstyle=\ttfamily\footnotesize,
    breakatwhitespace=false,         
    breaklines=true,                 
    captionpos=b,                    
    keepspaces=true,                 
    numbers=left,                    
    numbersep=5pt,                  
    showspaces=false,                
    showstringspaces=false,
    showtabs=false,                  
    tabsize=2
}
 
\usepackage{algorithm}  
\usepackage{algpseudocode}
\algblockdefx[Foreach]{Foreach}{EndForeach}[1]{\textbf{for each} #1 \textbf{do}}{\textbf{end for}}
\algrenewcommand{\algorithmiccomment}[1]{\hfill$\triangleright$ \emph{#1}}

\makeatletter 
\def\imod#1{\allowbreak\mkern8mu\-({\operator@font{}mod}\,{}\,#1)}  
\makeatother

\renewcommand{\vect}[1]{\ensuremath{\mathbf{#1}}}

\newtcolorbox{mybox}[2][]{%
  enhanced,
  colback=white,
  colframe=black,
  coltitle=white,
  fonttitle=\bfseries,
  boxrule=0.5mm,
  sharp corners,
  title=#2,
  #1
}
\let\proof\relax
\usepackage{amsthm}
\usepackage{stmaryrd}
\usepackage{xspace}

\theoremstyle{definition}

\newcounter{algcount}

\newcommand{\Psucc}{{\ensuremath{P_{\sf success}}}\xspace}

\newcommand{\hqcA}{{\textsf{hqc-128}}\xspace}
\newcommand{\hqcB}{\textsf{hqc-192}\xspace}
\newcommand{\hqcC}{\textsf{hqc-256}\xspace}
\newcommand{\PKEKG}{PKE.KeyGen\xspace}
\newcommand{\PKEE}{PKE.Encrypt\xspace}
\newcommand{\PKED}{PKE.Decrypt\xspace}

\newcommand{\KEME}{KEM.Encaps\xspace}
\newcommand{\KEMD}{KEM.Decaps\xspace}

\newcommand{\hashG}[0]{\mathcal{G}}
\newcommand{\hashK}{\ensuremath{\mathcal{K}}}
\newcommand{\Sample}{\ensuremath{\sample}}

\newcommand{\template}{\ensuremath{\mathcal{T}_{e,m}}}

\newcommand{\myindex}{\ensuremath{\mathcal{I}}\xspace}
\newcommand{\myindexA}{\ensuremath{\mathcal{I}_A}\xspace}
\newcommand{\myindexB}{\ensuremath{\mathcal{I}_B}\xspace}

\newcommand{\RM}[0]{{\sf Reed Muller}\xspace}
\newcommand{\RS}[0]{{\sf Reed Solomon}\xspace}
\newcommand{\ZT}[0]{{\sf Zero-Tester}\xspace}

\definecolor{iacrurlcolor}{HTML}{EC008C}
\usepackage[colorlinks=true,linkcolor=black!70!red,citecolor=black!70!green,urlcolor=iacrurlcolor]{hyperref}

\makeatletter
\def\email{\@ifnextchar[{\IACR@@email}{\IACR@email}}
\def\IACR@@email[#1]#2{\href{mailto:#1}{\nolinkurl{#2}}}
\def\IACR@email#1{\href{mailto:#1}{\nolinkurl{#1}}}
\makeatother

\author{Haiyue Dong \and Qian Guo}
\institute{
 Lund University, Lund, Sweden\\
 \href{mailto:haiyue.dong@eit.lth.se,qian.guo@eit.lth.se}{\textcolor{iacrurlcolor}{\texttt{\char123 haiyue.dong,qian.guo\char125 @eit.lth.se}}}
}

\title[OT-PCA: New KR-PC-SCAs on HQC with Offline Templates]{OT-PCA: New Key-Recovery Plaintext-Checking Oracle Based Side-Channel Attacks on HQC with Offline Templates}

\fancypagestyle{title}{%
  \fancyhf{}%
}

\begin{document}

\maketitle

\begin{abstract}
In this paper, we introduce OT-PCA, a novel approach for conducting Plaintext-Checking (PC) oracle based side-channel attacks, specifically designed for Hamming Quasi-Cyclic (HQC). By calling the publicly accessible HQC decoder, we build offline templates that enable efficient extraction of soft information for hundreds of secret positions with just a single PC oracle call. Our method addresses critical challenges in optimizing key-related information extraction, including maximizing decryption output entropy and ensuring error pattern independence, through the use of genetic-style algorithms.

Extensive simulations demonstrate that our new attack method significantly reduces the required number of oracle calls, achieving a 2.4-fold decrease for \hqcA and even greater reductions for \hqcB and \hqcC compared to current state-of-the-art methods. Notably, the attack shows strong resilience against inaccuracy in the PC oracle---when the oracle accuracy decreases to 95\%, the reduction factor in oracle call requirements increases to 7.6 for \hqcA.

Lastly, a real-world evaluation conducted using power analysis on a platform with an ARM Cortex-M4 microcontroller validates the practical applicability and effectiveness of our approach.

\keywords{Code-based cryptography \and NIST post-quantum cryptography standardization \and HQC \and Side-channel attacks \and KEM}
  
\end{abstract}

\section{Introduction}

Post-Quantum Cryptography (PQC) has rapidly become an essential area in modern cybersecurity, aimed at mitigating the significant threats posed by quantum computers to existing cryptographic infrastructures. These infrastructures largely depend on the computational challenges associated with factoring and discrete logarithms—both of which are susceptible to quantum attacks. In response, the National Institute of Standards and Technology (NIST) initiated a project in 2016 to identify and standardize quantum-resistant public-key cryptographic algorithms, focusing primarily on Key Encapsulation Mechanisms (KEM) and digital signatures. As part of this initiative, the lattice-based CRYSTALS-Kyber has been adopted as the primary KEM standard~\cite{NISTPQC:CRYSTALS-KYBER22}. Nevertheless, NIST continues to explore a diverse array of algorithms to ensure a secure and resilient cryptographic infrastructure. This includes evaluating code-based KEMs such as Classic McEliece~\cite{NISTPQC-R4:ClassicMcEliece22}, HQC~\cite{NISTPQC-R4:HQC22}, and BIKE~\cite{NISTPQC-R4:BIKE22}, as well as the isogeny-based SIKE~\cite{NISTPQC-R4:SIKE22}, which has been compromised. 

Among these, Hamming Quasi-Cyclic (HQC) emerges as a particularly promising candidate. NIST reports~\cite{915326} have highlighted that both BIKE and HQC, built on structured codes, are apt for general-purpose KEM applications not dependent on lattice structures.  NIST has indicated that it may choose one of these two for standardization after the fourth round of evaluation. Given this context, it is crucial to thoroughly assess HQC's security credentials and develop robust implementation strategies for various real-world platforms to ensure its effectiveness and reliability in a post-quantum world.

During the initial rounds of NIST’s evaluation, the implementation security of HQC has been extensively studied, as documented in significant research efforts~\cite{EPRINT:WBBGM19,SAC:PaiTer19,EC:BDHTV19,C:GuoJohNil20,AC:XIUTH21,DBLP:conf/pqcrypto/SchambergerHRWS22,TCHES:GHJLNS22,TCHES:HSCGJ23,DBLP:conf/pqcrypto/GoyLG22,AC:GNNJ23,EPRINT:PRJB23,DBLP:journals/tches/GoyMGL24,USS:SchroderGG24}. Among the various threats, Side-Channel Attacks (SCA) have emerged as particularly critical, necessitating thorough investigations to ensure secure implementations. These attacks exploit indirect data, such as timing differences, power consumption, or electromagnetic leaks, to extract sensitive information.

This paper focuses on a pivotal class of side-channel attacks in post-quantum cryptography, namely key-recovery Plaintext-Checking (PC) oracle based side-channel attacks targeting HQC. The objective of these attacks is to recover the security key of HQC by constructing a PC oracle through side-channel measurements. This PC oracle is capable of determining whether a specific ciphertext \( c \) corresponds to a predetermined message \( m \), thus facilitating the detection of decryption failures. Such an oracle can bypass Chosen Ciphertext Attack (CCA) secure transformations like the Fujisaki-Okamoto (FO) transformation~\cite{C:FujOka99}, compromising the security of Indistinguishability under Chosen Ciphertext Attack (IND-CCA) secure KEMs. 

These attacks are highly effective, exploiting a range of side-channel vulnerabilities as shown in studies on timing attacks~\cite{C:GuoJohNil20, TCHES:GHJLNS22}, power and electromagnetic (EM) attacks~\cite{TCHES:RRCB20, TCHES:UXTITH22, DBLP:conf/pqcrypto/SchambergerHRWS22}, and micro-architecture attacks~\cite{USENIX:WPHSFK22, gofetch, USS:SchroderGG24}, including cache timing attacks~\cite{TCHES:HSCGJ23}. They can leverage both substantial and minimal leakages, from strong power/EM signals~\cite{TCHES:UXTITH22} to subtle timing variations in high-performance CPUs~\cite{USS:SchroderGG24}.

Beyond identifying new sources of side-channel leakages for constructing a PC oracle, a primary research challenge lies in enhancing attack efficiency by minimizing the number of required PC oracle calls. Each oracle call translates to one or several side-channel measurements, which often become bottlenecks due to real-time interaction limitations with the target system. These limitations are imposed by system constraints, security measures, or other operational factors.

Earlier versions of HQC, prior to the third round of the NIST PQC competition, utilized a tensor product code that combined repetition codes with BCH codes. Certain parameter sets of these versions were susceptible to decryption failure attacks, as noted in~\cite{AC:GuoJoh20}. From the October 2020 release onward, HQC has shifted to using concatenated codes that integrate \RS~(RS) and \RM~(RM) codes. This shift introduces new challenges for PC oracle based attacks because the decoding properties of \RS and \RM codes are more complex than those of tensor product codes.

Recent advancements~\cite{TCHES:GHJLNS22, DBLP:conf/pqcrypto/SchambergerHRWS22, TCHES:HSCGJ23, AC:GNNJ23, USS:SchroderGG24} have significantly reduced the number of required PC oracle calls for the latest version of HQC. From the initial report at CHES 2022~\cite{TCHES:GHJLNS22}, which required over \numprint{800000} calls for \hqcA, subsequent works have progressively lowered this figure: to nine thousand at ASIACRYPT 2023~\cite{AC:GNNJ23}, and down to one thousand at USENIX Security 2024~\cite{USS:SchroderGG24}. These reductions mark substantial progress.  Yet, there remains the question of whether further reductions are feasible. Additionally, this research strives to develop algorithms optimized for practical applications, particularly in settings where constructing a perfectly accurate PC oracle is challenging, such as in complex environments or when robust security measures are in place.

\paragraph{Contributions}
We introduce OT-PCA, a novel approach to key-recovery PC oracle based side-channel attacks on HQC. 
We summarize the key contributions as follows.

 Our initial conceptual contribution is \myred{a novel method for efficiently extracting soft information representing the likelihood of entries in an HQC key block being non-zero. This method leverages the publicly available decoder to construct ``offline templates'' (OTs). Unlike traditional template attacks targeting specific devices, our OTs exploit the interaction in the decoding process between predefined error patterns and sparse vectors sampled from HQC's secret key distribution}.  Specifically, adding a sampled sparse vector to a given error pattern may cause bit flips in the latter, changing the likelihood of decryption successes or failures. \myred{By summarizing the sparse vectors leading to each decryption result, we empirically estimate position-wise probabilities, representing the probability of each position in the sparse vector being non-zero conditional on the decryption result.} During the online attack phase, when the decryption result of a new unknown sparse vector---the key block---is observed, we can thus infer the probability of each of the positions within the key block being non-zero using the \myred{estimated position-wise empirical probabilities}, i.e., offline templates.

This method enables the extraction of substantial soft information for hundreds of secret entries with a single oracle call, significantly enhancing the efficiency of PC oracle based side-channel attacks on HQC for full key recovery.

Our second major contribution tackles the challenge of identifying error patterns for template construction that maximize information extraction during subsequent attack stages. The most critical  element of this process is the first optimization problem---finding  error patterns that maximize the entropy of the decryption outcomes. By ensuring that each decryption output---success or failure---is approximately equally probable, our method effectively enhances the information gain from each future decryption instance.
In the paper, we advance our methodology by employing a set of error patterns instead of an individual pattern, thereby leveraging the collective information they provide. To ensure the patterns in the set are as independent as possible, we solve the second optimization problem---finding the set of error patterns that maximize the sum of pairwise Hamming distances. We develop simple genetic-style algorithms to efficiently solve these two problems, refining the error patterns used in constructing offline templates.

To summarize, the complete OT-PCA process begins with selecting several error patterns and constructing offline templates without calling the PC oracle. Subsequent online attack steps involve generating chosen ciphertexts, calling the PC oracle to obtain decryption outputs, and collecting secret information by combining these outputs with the offline templates. This process is further enhanced by a new, straightforward Information Set Decoding (ISD) algorithm that leverages soft information for efficient key recovery. 

Our simulation results demonstrate the high efficiency of this novel attack. A comparative analysis with previous methods using a perfect oracle (accuracy of 1) is shown in Table~\ref{table:comparison}, where our method significantly reduces the number of required PC oracle calls. For \hqcA, our approach achieves a 2.4-fold reduction in calls compared to the state-of-the-art~\cite{USS:SchroderGG24}. For \hqcB and \hqcC, the reductions are even more substantial, with improvements of 87 and 137 times, respectively, compared to~\cite{DBLP:conf/pqcrypto/SchambergerHRWS22}. These results mark major advancements over existing techniques, notably, the work~\cite{USS:SchroderGG24}, which struggles with~\hqcB.

Furthermore, with a less precise oracle (accuracy of 0.95), the improvement factor for \hqcA increases to 7.6 compared to the previous best approach~\cite{USS:SchroderGG24}. This demonstrates the robustness and efficiency of our attack under \myred{challenging conditions, such as exploiting subtle microarchitectural timing leakages~\cite{USS:SchroderGG24} or targeting masked implementations~\cite{TCHES:UXTITH22}}, where constructing a perfect PC oracle is difficult.

The OT-PCA method represents a significant advancement in PC oracle based attacks (PCA) on HQC, with potential applications beyond side-channel attacks to other domains such as misuse attacks~\cite{EC:BDHTV19,ACNS:HugVau20} and fault-injection attacks~\cite{AC:XIUTH21,DBLP:conf/indocrypt/HermelinkPP21}, whenever a PC oracle can be constructed.

 Lastly, we perform a real-world evaluation of our attack using power analysis on a platform with an ARM Cortex-M4 32-bit RISC core, targeting the PQClean implementation of \hqcA. The attacks were conducted across 120 different runs, each with a newly generated key. The outcomes from these real-world tests match our simulation results, demonstrating the practical effectiveness of our approach.

The source code is available at \url{https://github.com/ot-pca/ot-pca}.

 \begin{table}[!t]
\centering
\small
\begin{tabular}{lrrr}
    \toprule
    Work & \hqcA~ & \hqcB~ & \hqcC~\\
    \midrule
    GHJLNS22~\cite{TCHES:GHJLNS22} & $>$\numprint{800000} & --- & --- \\
SHRWS22~\cite{DBLP:conf/pqcrypto/SchambergerHRWS22} & \numprint{1152}$\times$46 & \numprint{1920}$\times$56 & \numprint{1920}$\times$90 \\
    HSCGJ23~\cite{TCHES:HSCGJ23} & $>$\numprint{50000} & --- & --- \\
    SCA-LDPC~\cite{AC:GNNJ23} & \numprint{9000} & --- & --- \\
    \ZT~\cite{USS:SchroderGG24} & \numprint{1094} & --- & ---\\
    Our work & \numprint{10}$\times$46 & 22$\times$56 & 14$\times$90\\
    \bottomrule
    \end{tabular}
\caption{Comparison of our attack with previous works: number of perfect PC oracle calls required for key recovery attacks against HQC. The symbol `---' indicates that the number is not provided in the corresponding work. \label{table:comparison}}
\end{table}

\paragraph{Organization}
The rest of the paper is organized as follows. Section~\ref{sec:pre} provides the necessary background on HQC and an overview of PC oracle based side-channel attacks on HQC, including their threat model and general approaches for recovering the secret key from the constructed PC oracle. Section~\ref{sec:newAttack} details the new attack methodology. In Section~\ref{sec:result}, we present simulation results for various accuracy levels of the assumed PC oracle and validate these findings on a real-world platform. This is followed by a discussion of the gains and potential extensions of the new attack methods \myred{in Section~\ref{sec:dis}}. Finally, in Section~\ref{sec:conclusion}, we conclude the paper and suggest future research directions.

\section{Preliminaries}
\label{sec:pre}

In this section, we provide the essential background on Hamming Quasi-Cyclic (HQC) codes and PC oracle based side-channel attacks on HQC. We begin by explaining the notations used throughout this paper.

\subsection{Mathematical Notations}
\label{subsection:math_notations}

Denote the binary finite field $\Ftwo$, where the operation $+$ is equivalent to the operation $-$.
 We work in HQC within the polynomial ring $\mathcal{R} = \Ftwo[X]/(X^n - 1)$, where $n$ is a positive integer. A polynomial $h$ within $\mathcal{R}$ can also be viewed as a row vector $\vect h$ of length $n$, in a vector space over $\Ftwo$, so we mix the notation if there is no ambiguity. 
The notion $\hamw{h}$ indicates the Hamming weight of $h$, i.e., the number of non-zero entries in the vector.
 The notation $[a..b]$ represents an index set starting from $a$ and ending at $b-1$, where $a$ and $b$ are two positive integers with $a < b$. Sampling operations are denoted as $\Sample(\mathcal{S})$ for uniform random sampling from a set $\mathcal{S}$, and $\Sample(\mathcal{R}, w)$ for sampling with the constraint that the elements have a Hamming weight of $w$.
The Bernoulli distribution, denoted by $\ber_p$, assigns the value 1 with probability $p$ and the value 0 with probability $1 - p$.

\subsection{Overview of Hamming Quasi-Cyclic (HQC)}
\label{subsection:hqc_overview}

Hamming Quasi-Cyclic (HQC)~\cite{NISTPQC-R4:HQC22} is a prominent code-based KEM that secures communication by relying on the difficulty of decoding random quasi-cyclic codes in the Hamming metric. Similar to other NIST post-quantum KEM proposals, HQC begins with an IND-CPA (Indistinguishability under Chosen Plaintext Attack) secure Public Key Encryption (PKE) scheme (\textsf{HQC.CPAPKE}) and transitions to an IND-CCA secure KEM (\textsf{HQC.CCAKEM}) through a CCA transformation. In HQC, this transformation employs the Hofheinz-Hövelmanns-Kiltz (HHK) method \cite{TCC:HofHovKil17}.

\begin{table}[t]
\centering
\small
\caption{HQC parameter sets as defined in~\cite{NISTPQC-R4:HQC22}. The base {\RM~code} used is the first-order $[128, 8, 64]$ {\RM~code}.} \label{tab:hqcParameters}
\begin{tabular}{lccccccccccc}
  \toprule
  & \multicolumn{3}{c}{RS-S} & \multicolumn{3}{c}{Duplicated RM} &&& \\
  \cmidrule(lr){2-4}\cmidrule(lr){5-7}
  Instance & $n_1$ & $k_1$ & $d_{\text{RS}}$ & Mult. & $n_2$ & $d_{\text{RM}}$ & $n_1n_2$ &$n$ & $\omega$ & $\omega_r = \omega_e$ \\
  \midrule
  \hqcA & $46$ & $16$ & $31$ & $3$ & $384$ & $192$ & $17664$ & $17669$ & $66$ & $75$\\
  \hqcB & $56$ & $24$ & $33$ & $5$ & $640$ & $320$ & $35840$ & $35851$ & $100$ & $114$\\
  \hqcC & $90$ & $32$ & $49$ & $5$ & $640$ & $320$ & $57600$ & $57637$ & $131$ & $149$\\
  \bottomrule 
\end{tabular}
\end{table}

\subsubsection{The Public Key Encryption (PKE) Core in HQC}
\label{subsection:hqc_pke}

The PKE core in HQC consists of three procedures: \PKEKG for key generation, $\text{\PKEE}$ for encryption, and $\text{\PKED}$ for decryption.

\paragraph{Three Procedures in \textsf{HQC.CPAPKE}}
During the key generation phase (\PKEKG), the scheme samples three polynomials: $h \sample \mathcal{R}$, $x \sample (\mathcal{R}, \omega)$, and $y \sample (\mathcal{R}, \omega)$. Consequently, we know that $\hamw{x} = \hamw{y} = \omega$. These polynomials define the secret key as $(x, y)$ and the public key as $(h, s = x + h \cdot y)$.

The encryption ($\text{\PKEE}$) process begins by generating polynomials $r_1\sample (\mathcal{R}, \omega_r)$, $r_2\sample (\mathcal{R}, \omega_r)$, and $e\sample (\mathcal{R}, \omega_e)$. This sampling process can be made deterministic by initializing the pseudo-random number generator (PRNG) with a seed $\theta$.
These components are then used to construct the ciphertext $(u, v)$, where $u = r_1 + h \cdot r_2$ and $v = m\mathbf{G} + s \cdot r_2 + e$. 
The matrix $\mathbf{G}$ represents a generator matrix for the linear code $\mathcal{C}$. In HQC, $\mathcal{C}$ is specifically designed as a concatenated code that combines \RM~and \RS~codes. 

The decryption procedure (\PKED) utilizes a decoder $\mathcal{C}.\text{Decode}(v - u \cdot y)$. Given that $s = x + h \cdot y$, we have:
\begin{align*}
v - u \cdot y &= m\mathbf{G} + s \cdot r_2 + e - (r_1 + h \cdot r_2) \cdot y \\
&= m\mathbf{G} + e',
\end{align*}
where
\[
e' = x \cdot r_2 - r_1 \cdot y + e.
\]
The decoder is designed with a specific error-correcting capability. If the Hamming weight $\hamw{e'}$ is sufficiently small, the decoder can correct the error, successfully recovering the encrypted message $m$. However, if the Hamming weight exceeds this capability, the decoder may fail, resulting in a decryption error.

\paragraph{Concatenated Code Construction in HQC}
The latest HQC specification uses a concatenated \RM~code and \RS~code. This process encodes an input message $m$ into a codeword $m\mathbf{G}$, where $\mathbf{G}$ is a publicly known generator matrix. The matrix $\mathbf{G}$ is defined in the space $\mathbb{F}_2^{k_m \times n_1 n_2}$, with $k_m$ set to $8k_1$.

The encoding process proceeds as follows: First, the message $m \in \mathbb{F}_{2^8}^{k_1}$ is encoded into $m_1 \in \mathbb{F}_{2^8}^{n_1}$ using the outer $[n_1, k_1, d_{\text{RS}}]$ \RS~code. Each byte $m_{1,i}$, where $0 \leq i < n_1$, is then encoded by the inner duplicated \RM\ code into $\bar{m}_{1,i} \in \mathbb{F}_{2}^{n_2}$. The resulting codeword is $m\mathbf{G} = (\bar{m}_{1,0}, \ldots, \bar{m}_{1,n_1-1})$ with a dimension of $n_1n_2$. Since computations in HQC are performed in the ambient space $\mathbb{F}_2^n$, any excess bits beyond $n_1 n_2$ are truncated when necessary.

The input to the function $\mathcal{C}.\text{Decode}$ is the vector $V = v - u \cdot y$, where $V\in\mathbb{F}_2^{n_1n_2}$. As shown in the decoding process outlined in Figure~\ref{fig:decoding}, $V$ is segmented into $n_1$ distinct blocks, labeled as $V = (V_0, \dots, V_{n_1-1})$, with each block $V_i$ within $\mathbb{F}_2^{n_2}$ and defined as an ‘inner block’ for $0 \leq i < n_1$. Each of these blocks $V_i$ undergoes decoding via the internal duplicated \RM~decoder, resulting in outputs $c_i^{\text{RS}}$ each within $\mathbb{F}_2^{8}$ for indices $0 \leq i < n_1$. These outputs are subsequently concatenated to form a sequence of $8n_1$ bits, represented as $c^{\text{RS}} = (c_0^{\text{RS}}, \ldots, c_{n_1-1}^{\text{RS}})$, with each $c_i^{\text{RS}}$ termed an ‘internal codeword’. This sequence $c^{\text{RS}}$, now a noisy codeword of the external \RS~code, is decoded into $k_1$ elements within $\mathbb{F}_{256}$ and subsequently translated into a message of $k_1$ bytes.

\usetikzlibrary{matrix, arrows.meta, positioning}

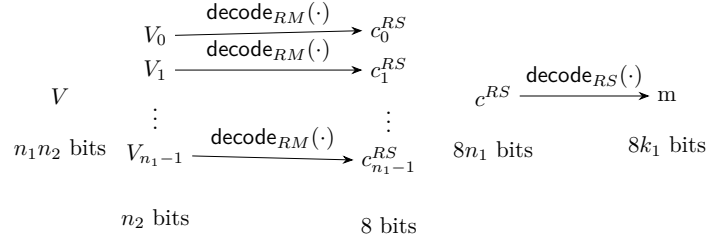
\begin{figure}[ht]
    \centering

    \resizebox{0.7\textwidth}{!}{
    \begin{tikzpicture}[>=Stealth, node distance=1cm]

\node (m) { m};
\node[below=0.3cm of m] (kbits) {$8k_1$ bits};
\node[left=2cm of m, baseline=(m.base)] (cRS) {$ c^{RS}$};
\node[below=0.3cm of cRS] (n1bits) {$8n_1$ bits};
\node[left=6cm of cRS, baseline=(cRS.base)] (cRMRS) {$V$};
\node[below=0.3cm of cRMRS] (n2bits) {$n_1n_2$ bits};

\draw[<-] (m) -- (cRS) node[midway, above] {${\sf decode}_{RS}(\cdot)$};

\matrix (matrix2) [matrix of math nodes, 
                  nodes={anchor=center}] at ([xshift=1.2cm]cRMRS.east)
{
  V_{0} \\
  V_{1} \\
  \vdots \\
  V_{n_1-1} \\
};

\matrix (matrix) [matrix of math nodes, 
                   nodes={anchor=center}] at ([xshift=-1.2cm]cRS.west)
{
   c_0^{RS} \\
  c_1^{RS} \\
  \vdots \\
   c_{n_1-1}^{RS} \\
};

\draw[<-] (matrix-1-1) -- (matrix2-1-1) node[midway, above] {${\sf decode}_{RM}(\cdot)$};
\draw[<-] (matrix-2-1) -- (matrix2-2-1) node[midway, above] {${\sf decode}_{RM}(\cdot)$};
\draw[<-] (matrix-4-1) -- (matrix2-4-1) node[midway, above] {${\sf decode}_{RM}(\cdot)$};

\node[below=0.3cm of matrix] {8 bits};
\node[below=0.3cm of matrix2] {$n_2$ bits};

\end{tikzpicture}
}
    \caption{Decoding of concatenated \RM~and \RS~codes}
    \label{fig:decoding}
\end{figure}

The three HQC parameter sets, \hqcA, \hqcB, and \hqcC, are presented in Table~\ref{tab:hqcParameters}.

\subsubsection{Key Encapsulation Mechanism (KEM) in HQC}
\label{subsection:hqc_kem}

\begin{figure}[ht]
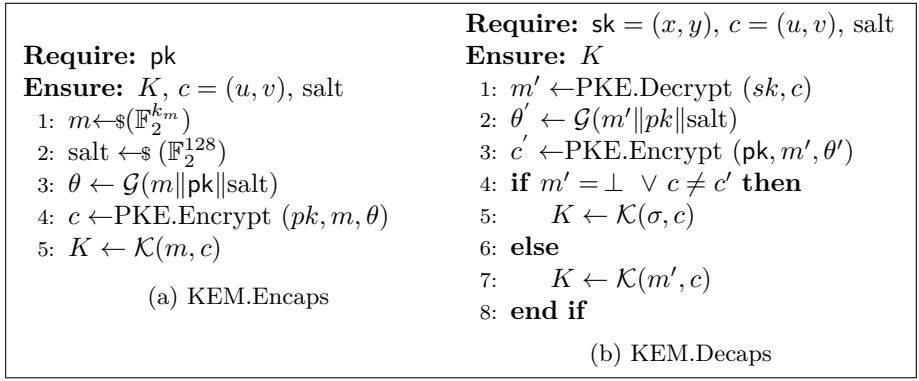

\centering
\fbox{
\begin{minipage}[h]{0.42\linewidth}
\begin{algorithmic}[1]
\Require $\pk$
\Ensure $K$, $c=(u,v)$, $\mathrm{salt}$
\State $m$\Sample$(\mathbb{F}_2^{k_m})$
\State $\mathrm{salt}\Sample(\Ftwo^{128})$
\State $\theta\gets\hashG(m \| \pk \| \mathrm{salt})$
\State $c\gets$\PKEE$(pk, m,\theta)$ 
\State $K\gets\hashK(m,c)$
\end{algorithmic}
\subcaption{\KEME}
\vspace*{3mm}
\end{minipage}
\begin{minipage}[h]{0.42\linewidth}
\begin{algorithmic}[1]
\Require $\sk=(x,y)$, $c=(u,v)$, $\mathrm{salt}$
\Ensure $K$
\State $m'\gets$\PKED$(sk,c)$
\State $\theta^{'}\gets\hashG(m'\|pk\|\mathrm{salt})$
\State $c^{'}\gets$\PKEE$(\pk, m',\theta')$
\If {$m' =\,\perp\ \lor\ c \neq c'$}
 \State $K\leftarrow\hashK(\sigma, c)$ 
\Else
\State $K\leftarrow\hashK(m', c)$
\EndIf
\end{algorithmic}
\subcaption{\KEMD}
\end{minipage}
}
\caption{{\sf HQC}.CCAKEM}\label{fig:KEM}

\end{figure}

The enhanced KEM variant of HQC, extending its PKE core, incorporates two crucial hash functions, $\hashG$ and $\hashK$. The process for constructing the IND-CCA secure KEM, denoted as \textsf{HQC.CCAKEM}, is illustrated in Figure~\ref{fig:KEM}. This construction includes a de-randomization of the encryption process via a seed $\theta$, which is derived by hashing the message $m$, the public key, and a random 128-bit salt. The decapsulation phase features a re-encryption step designed to validate the authenticity of the ciphertext.

\subsection{Overview of PC Oracle Based Side-Channel Attacks on HQC~\label{sec:overview}}

We describe the threat model of PC oracle based side-channel attacks (SCA) on the HQC KEM and provide an overview of the key recovery process.

\paragraph{The Threat Model}
This paper focuses on side-channel-assisted chosen-ciphertext attacks targeting the decapsulation procedure of the HQC KEM. In our model, an attacker can choose both the ciphertext $c$ and the message vector $m$, feeding them into the decapsulation algorithm on a specific device. The attacker leverages measurable side-channel leakages from this device.

Central to our threat model is the construction of a PC oracle, denoted as $\mathcal{O}_{\text{PC}}^{\rho}$. This oracle~\cite{EC:BDHTV19} utilizes side-channel leakage to assess whether $\text{\PKED}(sk, c) = m$. 
The oracle's accuracy, $\rho$, indicates the probability of a correct decision, with $1-\rho$ representing the probability of an error. 

The capability to construct such a PC oracle from side-channel leakage allows the attacker to exploit various types of side-channel leakages reported in the literature. This includes, but is not limited to, timing attacks~\cite{C:GuoJohNil20, TCHES:GHJLNS22}, power and electromagnetic (EM) attacks~\cite{TCHES:RRCB20, TCHES:UXTITH22, DBLP:conf/pqcrypto/SchambergerHRWS22, TCHES:SCZGJ23}, and micro-architecture attacks~\cite{USENIX:WPHSFK22, gofetch, USS:SchroderGG24} such as cache timing attacks~\cite{TCHES:HSCGJ23}. Such attacks demonstrate the vulnerability of various post-quantum KEMs, including those based on lattice, coding, and isogeny theories (cf.,~\cite{TCHES:UXTITH22}).

This study assumes a general PC oracle, allowing for broad applicability of our new attack methodology beyond the specific power leakages detailed in Section~\ref{sec:evalResult}. Consequently, our approach is adaptable to multiple forms of side-channel leakages once such a PC oracle can be constructed.

Moreover, the PC oracle based chosen ciphertext attacks are extensible to fault-injection attacks~\cite{AC:XIUTH21,DBLP:conf/indocrypt/HermelinkPP21}, including techniques like RowHammer~\cite{ACNS:MKBKV24}, enhancing the applicability of the newly developed method.

\paragraph{Overview of HQC Key Recovery Methods}
In the attack model, the attacker primarily gains information on decryption failures through a PC oracle constructed from side-channel leakages. To achieve full key recovery, the attacker must correlate this information with the hidden secret keys $y$ and $x$.

Research (e.g.,~\cite{TCHES:GHJLNS22, TCHES:HSCGJ23, AC:GNNJ23, USS:SchroderGG24}) reveals a strategic approach by attackers: the careful selection of polynomials $m$, $r_1$, $r_2$, and $e$ to craft ciphertexts $(u, v)$ with specific properties essential for extracting secret keys. Central to this approach is the dependency of decryption outcomes on the input to the decoder, $v - u \cdot y$. This approach ensures that decoding results are closely linked to the secret vector $y$, and indirectly to vector $x$, through the public parameter $h$ and the relationship $s = x + h \cdot y$.

These attack strategies assume that sufficient errors have already been introduced to push the outer \RS~decoder to its limit for correct decoding. Consequently, corrupting even a single additional \RM~block results in a decryption failure.
\section{The New Attack~\label{sec:newAttack}}

In this section, we present the new attack methodology, OT-PCA, which significantly reduces the number of oracle calls required in a PC oracle based attack against HQC. This reduction is achieved by constructing offline templates using the publicly available decoder for the concatenated code. We begin with a brief description of the key novel ideas behind this approach and then provide a detailed explanation of each step in the OT-PCA method.

\subsection{Novel Idea}
\label{sec:novel-approaches}

\paragraph{The Basic Setting}
Huang et al.~\cite{TCHES:HSCGJ23} pioneered a new strategy to reduce online PC oracle queries by leveraging the publicly available decoding function $\mathcal{C}.\text{Decode}(\cdot)$.
This approach was further refined in the study ~\cite{USS:SchroderGG24} presented at Usenix Security 2024. In these studies, the attacker strategically selects $r_1$ as a polynomial with a Hamming weight of one, specifically $X^k$, where $k$ is an integer, while setting $r_2$ to zero. This manipulation yields the ciphertext components $u = X^k$ and $v = m \mathbf{G} + e$. The polynomial fed to the decoding function $\mathcal{C}.\text{Decode}(\cdot)$ during the decapsulation is
\[
    v - u \cdot y = m \mathbf{G} + e - X^k \cdot y.
\]

Alternatively, by setting $r_2 = X^k$ and $r_1$ to zero, a different ciphertext configuration emerges: $u = X^k \cdot h$ and $v = m \mathbf{G} + s \cdot X^k + e$. The polynomial fed to the decoding function $\mathcal{C}.\text{Decode}(\cdot)$ during the decapsulation is
\[
    v - u \cdot y = m \mathbf{G} + s \cdot X^k + e - X^k \cdot h \cdot y = m \mathbf{G} + x \cdot X^k + e,
\]
assuming the relationship $s = x + h \cdot y$. 
With these chosen \( r_1 \) and \( r_2 \), the attacker can ascertain information about one length-\( n_2 \) block of the secret polynomials \( x \) and \( y \) by designing specific forms of \( e \). 
The chosen \( e \) should have \( \delta = \lfloor \frac{d_{\text{RS}}}{2} \rfloor \) blocks flipped to all ones, to ensure the outer \RS~decoding depends on the decoding output of the target \RM~block. 
The decoding output of the target \RM~block is sensitive to the values of certain positions in the secret polynomials \( x \) or \( y \), since they act as a perturbation of the vector \( e \) and together they are the input to the decoder.

Without loss of generality, the attacker can place the target length-$n_2$ block as the first block from position $0$ to $383$ by choosing a suitable value of $k$ for $X^k$. For instance, by choosing $k = -384$, the block with the index set $[384..768]$ is shifted to the first block starting with position $0$. A default setting for $m$ is $0$.

\paragraph{Constructing Offline Templates}
Building upon the basic settings proposed in~\cite{TCHES:HSCGJ23} and~\cite{USS:SchroderGG24}, we design a more efficient approach for selecting the first length-$n_2$ block of $e$ and extracting secret information from the constructed PC oracle. The key idea is that for a given vector $e$, the decryption result depends on the overall error of $e + X^k \cdot y$ (or $e + X^k \cdot x$). The vector $X^k \cdot y$ (or $X^k \cdot x$) is sparse and adding it to $e$ causes flips in $e$ at positions where the corresponding positions in $X^k \cdot y$ (or $X^k \cdot x$) are one. 
A flip in certain positions will increase the weight of the vector fed to the decoder, thereby increasing the likelihood of a failure. Conversely, a flip in other positions can decrease the weight and thus the likelihood of failure.
Therefore, conditional on the received decryption result, either `Success' or `Failure', the probability of a certain position in the first block of $X^k \cdot y$ (or $X^k \cdot x$) being one can be either higher or lower than the a-priori probability. This soft information can be obtained by calling the publicly available decoding function $\mathcal{C}.\text{Decode}(\cdot)$ and can be exploited for efficient full key recovery.

We begin by outlining the construction of the template given a selected error pattern $e$ and a message $m$, denoted as \template. Specifically, we generate \numprint{1000000} low-weight vectors $\mathbf{u}$ of length $n_2$, sampled according to their frequency in the secret key distribution. Initially, each vector element has a probability $p_a = \frac{\omega}{n}$ of being one, which for \hqcA, is approximately $3.74 \times 10^{-3}$. The vector sum $m\vect{G}_{\sf RM} + e + \mathbf{u}$, where $\vect{G}_{\sf RM}$ is a generator matrix of the \RM code, is then input into the \RM decoder to derive the decoding output, `Success' or `Failure', representing whether the decoder outputs the selected message $m$. 
To this output, we add a Boolean variable sampled from a Bernoulli distribution $\ber_{1-\rho}$ to emulate the behavior of the imprecise PC oracle $\mathcal{O}_{\text{PC}}^{\rho}$, which exhibits a decision error probability of $1-\rho$.

Given knowledge of the sampled vector $\mathbf{u}$, we can estimate the empirical conditional probability $\Pr(\mathbf{u}^i = 1 \mid \mathcal{O}_{\text{PC}}^{\rho})$ for $i \in [0..n_2]$.
This represents the likelihood that the $i$th position in the block is one, conditioned on the \RM decoding outcome. To determine this probability, we categorize $\mathbf{u}$ into two groups based on the decoding results. We then compute the frequency of the $i$th position being one within each group and normalize this by the total count of vectors in the respective group. 
This probability acts as an offline template \template, providing soft information crucial for further analysis.

\begin{figure}[ht]
\centering
\begin{subfigure}[b]{0.48\textwidth}
    \centering
    \includegraphics[width=\textwidth]{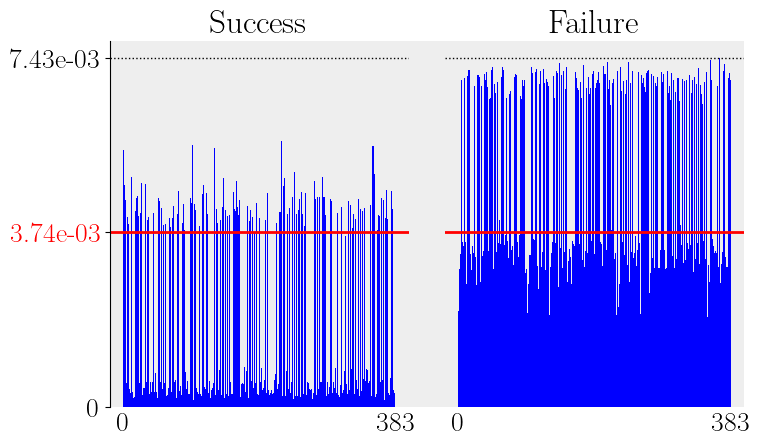}
    \caption{$\rho = 1$}
    \label{fig:subfig1}
\end{subfigure}
\hfill
\begin{subfigure}[b]{0.48\textwidth}
    \centering
    \includegraphics[width=\textwidth]{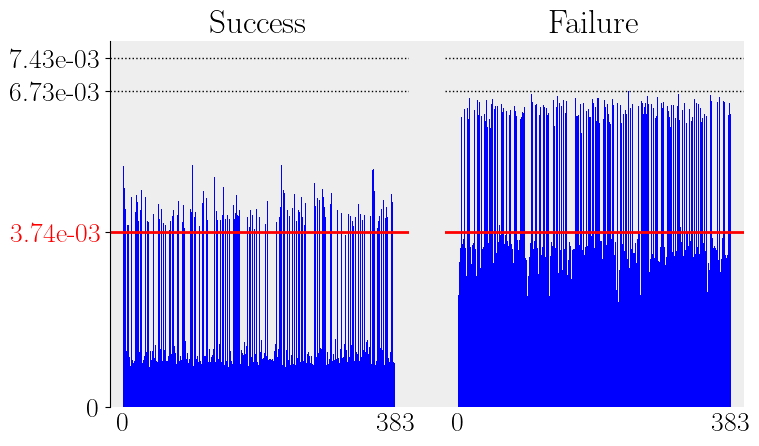}
    \caption{$\rho = 0.9$}
    \label{fig:subfig2}
\end{subfigure}
\caption{Constructed offline templates for \hqcA~using a chosen error vector $e_0$ with the message set to $0$. The x-axis represents the position within an \RM~block, and the y-axis represents the simulated empirical probability.}
\label{fig:OT}
\end{figure}

\paragraph{Example of Constructed Offline Templates for \hqcA}
Figure~\ref{fig:OT} presents two offline templates for \hqcA, derived from a specific error vector $e_0$ with the message set to $0$. Figure~\ref{fig:OT}a illustrates the case using a perfect oracle ($\rho = 1$), and Figure~\ref{fig:OT}b displays results from an oracle with 90\% accuracy ($\mathcal{O}_{\text{PC}}^{0.9}$). In both graphs, the red horizontal line indicates the a-priori probability $p_a \approx 3.74 \times 10^{-3}$, reflecting the initial likelihood of any position being one before decoding outputs are known.

The x-axis of each graph denotes the position within a \RM block, while the y-axis measures the simulated empirical conditional probability \myred{of a bit being 1 given the decryption outcome}. The decryption outcomes are labeled `Success' or `Failure'. With such decryption information, as shown in Figure~\ref{fig:OT}, many positions are significantly more likely to be zero or one. Notably, the comparison between the two panels highlights that a perfect oracle discloses more information, as evidenced by the greater deviations of the conditional probabilities from the a-priori probability $p_a$. This illustrates a reduced number of required PC oracle calls when utilizing a more accurate PC oracle.

\paragraph{Maximizing Extracted Information by Selecting Appropriate Error Patterns}
In the previous example, we demonstrated that the empirical conditional probability $\Pr(\mathbf{u}^i = 1 \mid \mathcal{O}_{\text{PC}}^{\rho})$ for $i \in [0..n_2]$ diverges from the a-priori probability $p_a$ once decryption results are disclosed. The attacker selects a suitable message $m$, compiles a list of $t$ error patterns $\mathcal{E} = \{e_1, e_2, \ldots, e_t\}$, collects the corresponding decryption outputs $\vect{o} = (o_1, o_2, \ldots, o_t)$, and performs a posteriori estimation by calculating:
\begin{equation}
\label{eq:1}
    \Pr(\mathbf{u}^i = 1 \mid \vect{o}) \myred{~\propto  } \prod_{j=1}^{t}\Pr(\mathbf{u}^i = 1 \mid o_j),
\end{equation}
assuming independence among decryption outputs. \myred{Since our objective is to rank these probabilities for the secret bits, we focus on calculating the right-hand side of Equation~\ref{eq:1}.}
The probability $\Pr(\mathbf{u}^i = 1 \mid o_j)$ can be read from the $t$ constructed offline templates $\mathcal{T}_{e_j,m}$, for $i \in [0..n_2]$ and $j \in [1..t+1]$.

A key challenge is determining which error patterns to choose in our attacks, which involves solving two optimization problems.
\begin{itemize}
 \item {\bf Maximizing Decryption Output Entropy:}
    Firstly, the PC oracle reveals a Boolean output that can be viewed as a binary random variable. 
    Our goal is to make this random variable as uniform as possible, achieving an entropy close to $1$. The binary entropy function $H(p_d)$ is defined as $H(p_d) = -p_d \log_2(p_d) - (1 - p_d) \log_2(1 - p_d)$, where $p_d$ is the probability of decryption failures. By maximizing the entropy, we ensure that the information obtained from \myred{one oracle call} is maximized, in accordance with information theory.
    \item {\bf Making the Errors Different: } Secondly, \myred{to maximize information gain from multiple oracle queries,} we aim to maximize the sum of Hamming distances:
    \[
        \sum_{1 \leq i < j \leq t} \hamw{e_i + e_j}.
    \]
\end{itemize}

To address the first optimization problem, we implement a heuristic algorithm inspired by mutation fuzzers. Our objective is to fine-tune an error pattern such that the \RM decoding result yields uniformly distributed outputs after introducing minor perturbations determined by the secret distribution of $y$.

We begin with an initial error pattern $e_{ini}$ of dimension $n_2$, which is randomly sampled to meet the condition:
\[
\lfloor \theta_0 \cdot n_2 \rfloor \leq \text{wt}(e_{ini}) \leq \lfloor \theta_1 \cdot n_2 \rfloor,
\]
where $\text{wt}(e_{ini})$ denotes the Hamming weight of $e_{ini}$, and $\theta_0$ and $\theta_1$ are parameters specific to the target HQC parameter sets. For \hqcA, we choose $\theta_0 = 0.42$ and $\theta_1 = 0.44$, while for \hqcB and \hqcC, we select $\theta_0 = 0.435$ and $\theta_1 = 0.460$. These parameters are determined experimentally to ensure that the algorithm starts with an effective pattern, allowing it to quickly reach a near-optimal solution.

Next, we employ a simple genetic-style algorithm to refine the error pattern. During the mutation phase, the error pattern is dynamically adjusted by flipping one or two bits, based on the evaluation function $\mathcal{F}(\cdot)$, which calculates the binary entropy function of the output distribution from the \RM decoder. The input to the \RM decoder is the current binary error pattern $\hat{e}$ XORed with a perturbation vector that follows the HQC secret key distribution.

If the entropy is below a specified threshold, two bits are flipped to increase search diversity; otherwise, a single bit is flipped for finer adjustments. Each mutated pattern at iteration $k$ is denoted as $\hat{e}^{(k)}$. The optimization criterion is applied as follows:
\[
\hat{e}^{(k+1)} = \argmax_{\hat{e}' \in \text{Mutate}(\hat{e}^{(k)})} \mathcal{F}(\hat{e}'),
\]
where $\text{Mutate}(\cdot)$ represents the bit-flipping operation.

The algorithm iteratively updates and evaluates each generation, refining the error pattern to maximize fitness until reaching the predetermined number of generations. This process is executed concurrently across hundreds of threads, yielding a long list of optimized $e_i$. After this procedure, we retain a list of the patterns with binary entropy greater than a threshold, which is set to be $0.999$ in our experiments. 

The second optimization goal is to identify a set of $t$ vectors from the list that maximize the sum of pairwise Hamming distances. A genetic algorithm with multiple random initializations is applied. Specifically, the algorithm starts with a randomly selected set of $t$ vectors $e_i$ and employs a genetic algorithm that iteratively evolves this set. In each generation, mutations are produced by randomly substituting one vector from the set with a non-duplicate from the list. The set demonstrating the highest fitness is selected as the parent set for the subsequent generation. This process continues until a predefined number of generations is reached, ultimately yielding the optimal set of vectors. To robustly explore the solution space and avoid local optima, we repeat the entire process across numerous random initializations, each independently seeking the optimal combination of vectors. The final output is the set of $t$ vectors that demonstrates the maximum sum of pairwise Hamming distances observed across all iterations.

While the second optimization goal is maintained, it provides only marginal gains. This minimal impact is due to the fact that the two randomly selected elements, $e_i$ and $e_j$, already possess a relatively large Hamming distance following the first optimization. Therefore, the additional benefits from this subsequent optimization step are minor.

The offline optimization phase is independent of a specific attack instance. Notably, if these pre-computed error vectors were publicly accessible, they could facilitate the launch of an OT-PCA attack by any adversary.

\subsection{Online Attack}
\begin{figure}[ht!]
\includestandalone[scale=0.85,mode=buildnew,width=0.85\columnwidth]{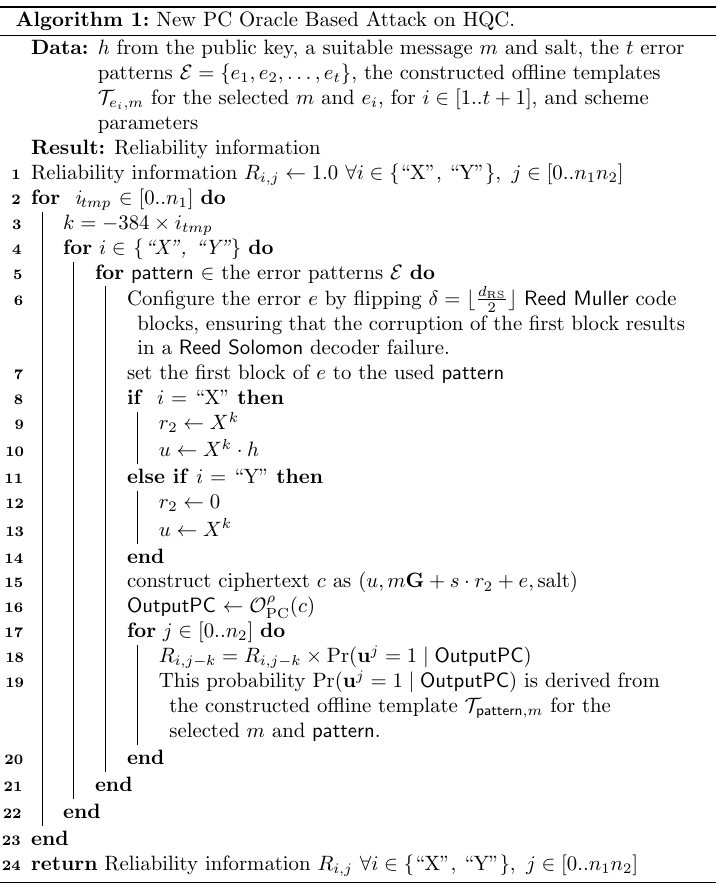}
\caption{The online phase of the new OT-PCA attack on HQC.}
\label{fig:online}
\end{figure}

Following the ciphertext selection method described in Section~\ref{sec:novel-approaches}, we feed the selected ciphertexts to the constructed PC oracle to obtain decryption results.
The pseudocodes of the online attack phase for this new PC oracle based attack are provided in Figure~\ref{fig:online}. 

In this attack, we define an integer $k = -384 \times i$ for $i \in [0..n_1]$ to shift each of the \RM blocks to the first block. For each shift, we obtain $t$ measurements for the $n_2$ secret entries in each block. After $n_1$ shifts, we gather soft information for the first $n_1\cdot n_2$ secret entries of $x$ and $y$, respectively.
By concatenating these $2 n_1 \cdot n_2$ secret entries, we form a new vector $\vect{z}$. Consequently, the total number of required online PC oracle calls is $2 \cdot t \cdot n_1$.

Using the constructed offline template, we rank the conditional probability $\Pr(\mathbf{z}_i = 1 \mid \vect{o})$ by computing the product
\(\prod_{i=1}^{t}\Pr(\mathbf{z}_i = 1 \mid o_i),
\)
for $0 \leq i < 2n_1n_2$. This product provides soft reliability information about the secret entries. A lower value of this product indicates a higher likelihood that the corresponding secret entry is 0, based on the decryption outputs from the PC oracle.

\subsection{Full Key Recovery from ISD with Soft Information}
\label{sec:postPro}

During the online attack phase, we collect the reliability information for $2n_1n_2$ positions of the secret $(x, y)$. This enables us to select the most reliable $n + l$ positions for performing Information Set Decoding (ISD) with soft information, where $l$ is a small positive integer. We ensure that the allowable post-processing cost is comparable to Gaussian elimination, facilitating a fair comparison with~\cite{USS:SchroderGG24}. We propose a modified Stern-style algorithm with simplified complexity estimation. Alternatively, for improved performance, one might consider adopting a more advanced ISD algorithm with soft information, such as the SoftStern algorithm described in \cite{guo2019some}.

For an \( n \times 2n \) parity-check matrix, the cost of Gaussian elimination, denoted as \(C_{\text{Gauss}}\), is \(\ordo{n^3}\). According to an analysis in~\cite{meurer2013coding}, the concrete cost of a simple implementation can be estimated as \(2^{43}\), \(2^{46}\), and \(2^{48}\) for \hqcA, \hqcB, and \hqcC, respectively.

\paragraph{Quasi-Standard Form}
We describe a new but simple approach to exploit the soft information to accelerate the ISD algorithm. Suppose we have a system of linear equations
\begin{equation}
\label{eq:keyEq}
\begin{bmatrix} I_n & \mathrm{rot}(h) \end{bmatrix} \begin{bmatrix} x \\ y \end{bmatrix} = s,
\end{equation}
which is obtained from the public key $s = x + h \cdot y$, where $\mathrm{rot}(h)$ represents the circulant matrix induced by $h$. In this matrix, $h$ is placed as the first column, and each subsequent column is a cyclic shift of the previous one.

Finiasz and Sendrier in \cite{AC:FinSen09} first proposed a general ISD framework---choosing a permutation of the positions and transforming the matrix $\begin{bmatrix} I_n & \mathrm{rot}(h) \end{bmatrix}$ into a quasi-standard form
\[
\widetilde{\mathbf{H}} := \begin{pmatrix}
    \mathbf{Q}' & \mathbf{0} \\
    \mathbf{Q}'' & \mathbf{I}_{n-\ell}
\end{pmatrix},
\]
where $\mathbf{Q}'$ is a $\ell \times (n + \ell)$ matrix, $\mathbf{Q}''$ is a $(n - \ell) \times (n + \ell)$ matrix, and $\mathbf{0}$ is a $\ell \times (n - \ell)$ zero matrix.

\paragraph{New ISD Exploiting Soft Information}
Instead of randomly choosing $n + \ell$ positions to form the matrix $\mathbf{Q}'$ as in standard ISD algorithms~\cite{meurer2013coding}, we leverage the soft information obtained from the online decryption oracle calls. Specifically, we reorder the index set $[0..2n_1n_2]$ in increasing order according to the probability of the secret entries being 1 and select the top $n + \ell$ positions with the smallest probabilities.

Given that the top positions can have very low probabilities of being 1, we introduce two algorithm parameters, $T_0$ and $T_1$. We draw $T_1$ positions from the $T_0$ positions with the smallest probabilities of being 1, and assume that these $T_1$ positions are all zero. These positions are then removed from $\widetilde{\mathbf{H}}$. After removing the selected $T_1$ positions from the $n + \ell$ positions with the smallest probabilities, we denote the resulting index set as \myindex.

Next, we adopt a collision procedure that splits the index set \myindex into two parts of equal size $\frac{n+\ell - T_1}{2}$, denoted by \myindexA and \myindexB, respectively. The parameter $\ell$ is picked to ensure that $n+\ell - T_1$ is an even number. We then build two lists to enumerate bit patterns of length $\frac{n+\ell - T_1}{2}$ with Hamming weight bounded by $w_0$. We include the syndrome information in one list to find collisions that meet the partial syndrome of $\ell$ positions in the Finiasz-Sendrier framework. To ensure that the overall complexity of the new ISD is not much larger than one Gaussian elimination, we only consider $w_0$ values of 2 or 3. The complexity of generating the two lists is thus
\begin{equation}
C_{\text{list}} = 2 w_0 \cdot \ell \cdot \binom{\frac{n+\ell - T_1}{2}}{w_0}.
\end{equation}

The collision procedure will output a list of potential candidates that need to be verified to ensure that Equation~\ref{eq:keyEq} is satisfied. We choose $\ell$ to be sufficiently large, making the cost of this verification step negligible.

\paragraph{Summary}
Our method involves performing one Gaussian elimination process and $n_{iter}$ inner rounds, with $n_{iter}$ set to 16 by default. The dominant part of the complexity is $\max\{C_{\text{Gauss}}, n_{iter} \cdot C_{\text{list}}\}$. The concrete cost will be calculated in Section~\ref{sec:sim}, where experimental parameters determine the success probability. When $w_0$ is set to 3, the added cost is comparable to Gaussian elimination; when $w_0$ is set to 2, the added cost is significantly lower than one Gaussian elimination. 
For the $n + \ell$ positions with the smallest probabilities of being 1, the algorithm will succeed if, during one of the $n_{iter}$ inner rounds, both (1) the $T_1$ positions set to zero are error-free, and (2) no more than $w_0$ errors occur in the positions in the index sets \myindexA and \myindexB, respectively. We denote this success probability as $P_{\sf success}$. \myred{The suitable values for parameters $T_0$ and $T_1$ are determined empirically. }

\section{Results~\label{sec:result}}

In this section, we present the results of our attack methodology, obtained from extensive simulations and real-world evaluations. The results are divided into two subsections: the first details the success probability of the attack obtained through simulations, considering a general PC oracle constructed from side-channel measurements with varying levels of accuracy. The second describes the real-world evaluation using power traces from a platform equipped with an ARM Cortex-M4 core.

\subsection{Simulation Results~\label{sec:sim}}
We begin with the simulation results, assuming a general PC oracle that can be constructed from side-channel measurements. Extensive simulations were conducted to evaluate the performance of the new attack with \numprint{10000} random keys tested in each instance. These results are organized into two parts: one focusing on \hqcA, and the other on \hqcB and \hqcC. Given that \hqcA has been the primary focus in previous research, we conduct a detailed investigation of its attack performance across oracle accuracies of 1, 0.995, 0.95, and 0.9. This comprehensive analysis allows us to predict the performance of our methodology in various scenarios where an accurate PC oracle is difficult to obtain. For \hqcB and \hqcC, we limit our study to a PC oracle accuracy of 1. While this study primarily focuses on \hqcA, there are no inherent limitations preventing the application of our methodology to \hqcB and \hqcC with less accurate oracles.

\begin{figure}[t]
\centering
\includegraphics[width=0.8\textwidth]{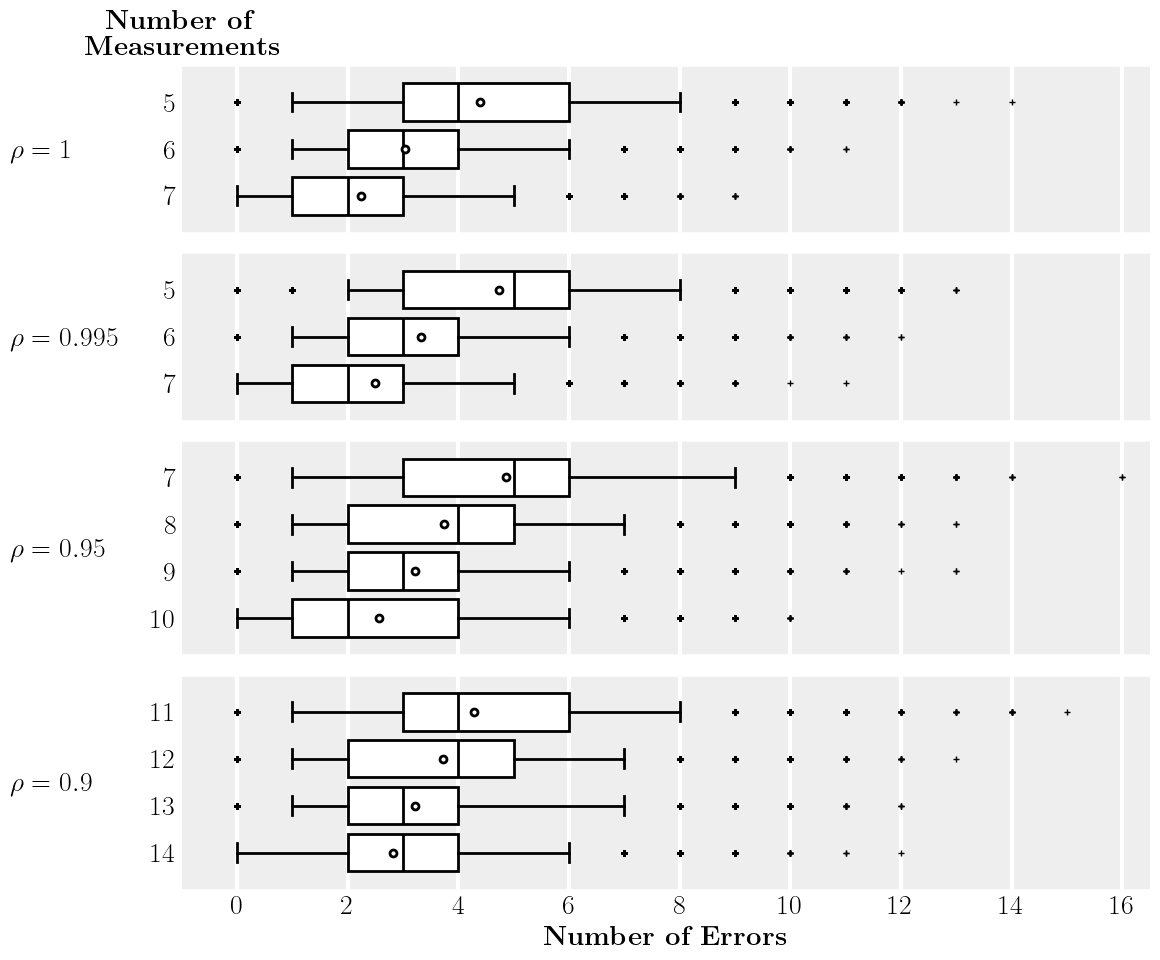}
\caption{Number of measurements vs. number of errors for \protect\hqcA with various oracle accuracies, based on \numprint{10000} random keys tested in each scenario.
The number of errors in the first $n$ positions of the reordered $\vect z$ vector decreases as the number of side-channel measurements per length-$n_2$ block increases.
The whiskers represent the 5th and 95th percentiles. The small circles denote the mean number of errors, which are close to the median values indicated by the central line in each box. }
\label{fig:measurementsVSerrorsHQCA}
\end{figure}

\subsubsection{Targeting \hqcA}
In our attack, we aim to shift the ones in the secret key towards the end of the vector after reordering based on the soft information obtained. Non-zero values within the first $n$ positions of this reordered sequence are considered errors. Figure~\ref{fig:measurementsVSerrorsHQCA} shows the simulated number of errors for different oracle accuracies $\rho$ across various numbers of PC oracle calls per length-$n_2$ block. Consequently, the total number of PC oracle calls needs to be multiplied by a factor of $2n_1$, which is $92$ for \hqcA.

It is evident that the number of simulated errors in the first $n$ positions of the reordered vector $\vect{z}$ decreases as the number of side-channel measurements increases. We observed in simulation that when the mean error numbers are no more than 5, we achieve a success rate \Psucc higher than 50\% when $w_0=3$.

\paragraph{Post-Processing Complexity vs. Success Rate}
In the post-processing phase for \hqcA, we apply our new ISD algorithm with parameters $\ell = 51$, $T_0 =$ \numprint{12000}, $T_1 =$ \numprint{10000}, and $n_{iter} =$ \numprint{16}. Consequently, the cost of $n_{iter} \cdot C_{\text{list}}$ is approximately $2^{45}$ for $w_0 = 3$, and $2^{34}$ for $w_0 = 2$. The list sizes for the two lists in the collision search are $2^{33}$ for $w_0 = 3$ and $2^{23}$ for $w_0 = 2$. The simulated success probability \Psucc is presented in Table~\ref{tab:SRHqc128}. We observe that the success probability increases drastically as the number of PC oracle calls per length-$n_2$ block, denoted by $t$, increases.

\myred{Although ideally templates should be tailored to specific oracle precisions, our experiments demonstrate the robustness of a single template constructed using a perfect oracle. This template exhibits consistent effectiveness across various accuracy levels with minimal performance degradation. For instance, when applied to oracles with $\rho = 0.9$, $\omega_0 = 3$, and $t$ ranging from 11 to 14, the success probabilities remained consistently high (65.95\%, 75.18\%, 82.97\%, and 87.02\% respectively), closely aligning with the results from Table~\ref{tab:SRHqc128}. }

\begin{table}[t]
\centering
\small
    \setlength\tabcolsep{0.7cm}
\caption{Simulated success rates \(\protect\Psucc\) for \protect\hqcA, illustrating the influence of oracle accuracy (\(\rho\)) and the number of PC oracle calls per length-\(n_2\) block (\(t\)). The parameter \(w_0\) is defined in Section~\protect\ref{sec:postPro}. \numprint{10000} random keys are tested in each scenario.}
\label{tab:SRHqc128}
\begin{tabular}{@{}lccc@{}}
\toprule
$\rho$ & $t$ &                                                              $w_0=2$ & $w_0=3$  \\                
\midrule
1                   & 5                                                                                  & 43.42\%               & 65.10\%              \\
                        & 6                                                                                  & 70.18\%               & 83.53\%              \\
                        & 7                                                                                  & 85.45\%               & 92.60\%              \\
                        &                                                                                    &                       &                      \\
0.995               & 5                                                                                  & 36.65\%               & 59.25\%              \\
                        & 6                                                                                  & 65.10\%               & 80.61\%              \\
                        & 7                                                                                  & 82.13\%               & 91.80\%              \\
                        &                                                                                    &                       &                      \\
0.95                & 7                                                                                  & 36.03\%               & 57.45\%              \\
                        & 8                                                                                  & 57.48\%               & 76.27\%              \\
                        & 9                                                                                  & 68.23\%               & 83.72\%              \\
                        & 10                                                                                 & 80.12\%               & 90.02\%              \\
                        &                                                                                    &                       &                      \\
0.9                 & 11                                                                                 & 48.44\%               & 67.58\%              \\
                        & 12                                                                                 & 58.79\%               & 76.40\%              \\
                        & 13                                                                                 & 68.05\%               & 83.37\%              \\
                        & 14                                                                                 & 75.60\%               & 87.55\%              \\ \bottomrule
\end{tabular}
\end{table}

\paragraph{Comparison with Previous Works}
Table~\ref{tab:medianoraclcalls} compares our results with the state-of-the-art as represented by SCA-LDPC~\cite{AC:GNNJ23} and \ZT~\cite{USS:SchroderGG24}, over a range of oracle accuracies $\rho$. Both studies report the median number of oracle calls, so for a fair comparison, we report the number of oracle calls required to achieve a success rate (\(\Psucc\)) greater than 50\%, indicating that at least half of the attack instances are successful. 

Our approach demonstrates remarkable improvements, consistently requiring fewer oracle calls across all tested accuracies, thereby showing substantial efficiencies. The improvement factor is particularly striking, reaching as high as 58.8 compared to SCA-LDPC and 7.6 against \ZT as the oracle accuracy decreases. 

Furthermore, our method’s performance advantage becomes more pronounced with decreasing oracle accuracy, showing its effectiveness in challenging scenarios. This robustness is essential for practical applications, especially in security-sensitive contexts where effective countermeasures are deployed and perfect oracles are rare.

\begin{table*}[ht]
    \centering
    \small
    \begin{tabular}{lrrrr}
    \toprule
    $\rho $ & $ 1$ & $0.995$ & $0.95$ & $0.9$ \\
    \midrule
    SCA-LDPC~\cite{AC:GNNJ23} & \numprint{9000}  & \numprint{18000} & \numprint{35250} & \numprint{59500}  \\
    \ZT~\cite{USS:SchroderGG24} & \numprint{1094}  & \numprint{2246} & \numprint{4922} & \numprint{6951}  \\
    \midrule
    Our Work & \numprint{5}$\times$92  & \numprint{5}$\times$92 & \numprint{7}$\times$92 & \numprint{11}$\times$92 \\
     Ratio A & 19.6  & 39.1 & 54.7 & 58.8 \\
    Ratio B & 2.4  & 4.9 & 7.6 & 6.9\\
    \bottomrule
    \end{tabular}
    \caption{Comparison of median oracle calls. For our work, we report the required number of oracle calls needed to achieve a success rate greater than 50\%, indicating that at least half of the attack instances are successful. Our results are compared with SCA-LDPC~\cite{AC:GNNJ23} and \ZT~\cite{USS:SchroderGG24} for various oracle accuracies $\rho$. Ratio A shows the ratio of oracle calls for SCA-LDPC to our method, and Ratio B shows the ratio for \ZT to our method, highlighting our approach's efficiency.}
    \label{tab:medianoraclcalls}
\end{table*}

\begin{figure}[ht]
\centering
\small
\includegraphics[width=0.8\textwidth]{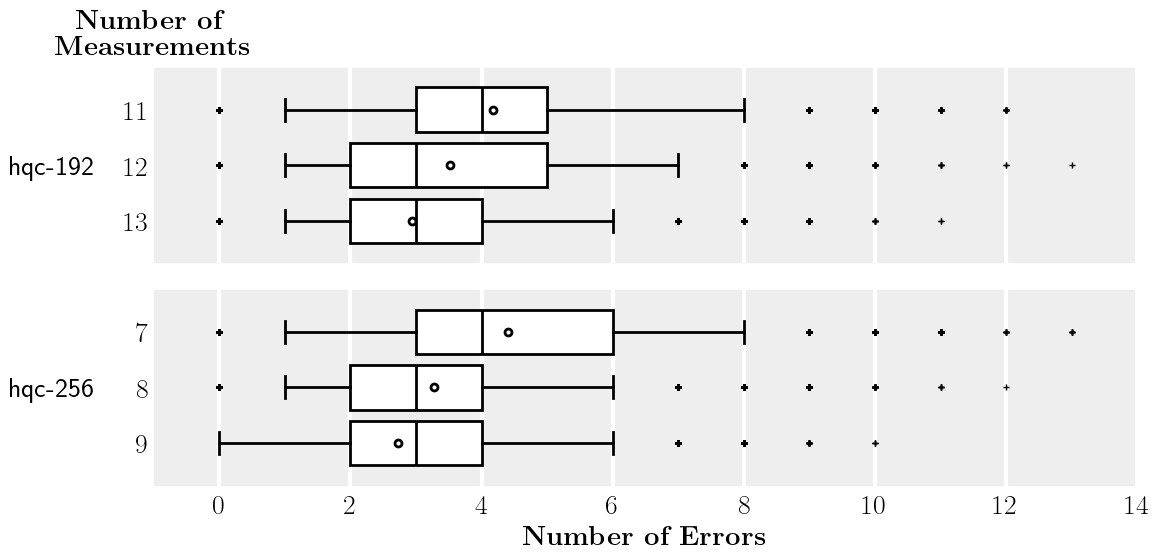}
\caption{Number of measurements vs. number of errors for \textsf{hqc-192}~and \textsf{hqc-256}~with a perfect oracle, based on \numprint{10000} random keys tested in each scenario.
The number of errors in the first $n$ positions of the reordered $\vect z$ vector decreases as the number of side-channel measurements per length-$n_2$ block increases.
The whiskers represent the 5th and 95th percentiles. The small circles denote the mean number of errors, which are close to the median values indicated by the central line in each box. }
\label{fig:measurementsVSerrors192}
\end{figure}

\subsubsection{Targeting \hqcB and \hqcC}
Figure~\ref{fig:measurementsVSerrors192} shows the number of errors from simulations for \hqcB and \hqcC. This analysis varies the number of PC oracle calls per length-$n_2$ block, assuming a perfect PC oracle has been constructed. For these scenarios, the total number of PC oracle calls also needs to be adjusted by a factor of $2n_1$, which translates to $112$ for \hqcB and $180$ for \hqcC. Similar to \hqcA, the number of simulated errors in the initial $n$ positions of the reordered secret vector decreases steadily as the number of measurements increases.

\begin{table}[t]
\centering
\small
    \setlength\tabcolsep{0.7cm}
\caption{Simulated success rates \(\protect\Psucc\) for \protect\hqcB and \protect\hqcC, assuming a perfect PC oracle.  The number of PC oracle calls per length-\(n_2\) block is denoted by \(t\). Parameter \(w_0\) is detailed in Section~\protect\ref{sec:postPro}. \numprint{10000} random keys are tested in each scenario.}
\label{tab:SRHqcBC}
\begin{tabular}{@{}lccc@{}}
\toprule
 & $t$ &                                                              $w_0=2$ & $w_0=3$  \\                
\midrule
\hqcB & 11                                                                                & 47.90\%               & 68.38\%              \\
        & 12                                                                                & 61.93\%               & 78.06\%              \\
        & 13                                                                                & 72.47\%               & 84.81\%              \\ 
                                &                                                                                    &                       &                      \\
\hqcC & 7                                                                                 & 49.45\%               & 73.01\%              \\
        & 8                                                                                 & 70.93\%               & 87.51\%              \\
        & 9                                                                                 & 81.37\%               & 92.47\%              \\ \hline
\end{tabular}
\label{tab:hqcBC}
\end{table}

\paragraph{Post-Processing Complexity vs. Success Rate}
During the post-processing phase for \hqcB, we implement our novel ISD algorithm with parameters $\ell = 61$, $T_0 = \numprint{24000}$, $T_1 = \numprint{20000}$, and $n_{iter} = \numprint{16}$. For \hqcC, the parameters are adjusted to $\ell = 61$, $T_0 = \numprint{40000}$, $T_1 = \numprint{30000}$, and $n_{iter} = \numprint{16}$. This phase reveals that the added computational cost for \hqcB, calculated as $n_{iter} \cdot C_{\text{list}}$, reaches approximately $2^{49}$ when $w_0 = 3$, and drops to $2^{37}$ with $w_0 = 2$. The sizes of the two lists used in the collision search are $2^{36}$ and $2^{25}$ for $w_0 = 3$ and $w_0 = 2$, respectively. In contrast, for \hqcC, these costs increase to about $2^{51}$ for $w_0 = 3$ and $2^{38}$ for $w_0 = 2$, with the corresponding list sizes at $2^{39}$ and $2^{27}$. Tabel~\ref{tab:hqcBC} showcases the simulated success rates, \Psucc. Mirroring trends observed with \hqcA, there is a marked increase in success probability as the number of PC oracle calls per length-$n_2$ block ($t$) increases.

\paragraph{Comparison with Previous Works}
In the context of key-recovery PC oracle based side-channel attacks against \hqcB and \hqcC, the fewest number of required PC oracle calls were documented by Schamberger et al.~\cite{DBLP:conf/pqcrypto/SchambergerHRWS22}, which is \(1920 \times 56\) for \hqcB and \(1920 \times 90\) for \hqcC. Our OT-PCA method achieves significant reductions with only minor post-processing costs---\(22 \times 56\) for \hqcB and \(14 \times 90\) for \hqcC. These reductions translate to remarkable factors of \(87\) for \hqcB and \(137\) for \hqcC, demonstrating a major improvement in efficiency.

Additionally, while Schröder et al. in~\cite{USS:SchroderGG24} have noted challenges in applying the \ZT method to \hqcB, our methodology, as discussed in Section~\ref{sec:MoreZT}, provides robust performance for \hqcB.

\subsection{Real-World Validation~\label{sec:evalResult}}
We demonstrate the feasibility and effectiveness of our newly proposed attack methodology by conducting a power analysis attack on a real-world platform.

\paragraph{Experimental Setup}
Our experimental setup consists of the ChipWhisperer Husky, the CW313 base-board, and an ATSAM4S2A microcontroller target with an ARM Cortex-M4 32-bit RISC core. The operating frequency is set to 24 MHz, with a sampling rate of 24 MS/sec.
We target the PQClean implementation of \hqcA.  Specifically, we incorporate trigger functions to capture the waveform during the execution of the {\sf PQCLEAN\_HQC128\_CLEAN\_code\_encode} function, which is a crucial part in the re-encryption phase of HQC decapsulation \KEMD. This function calculates the encoding $m'\vect{G}$ for the decrypted message $m'$.
The implementation is compiled using {\sf arm-none-eabi-gcc} with the highest optimization level  {\sf -O3}. 

\begin{figure}[ht]
    \centering
    \small
    \begin{subfigure}[b]{0.85\textwidth}
        \centering
        \includegraphics[width=\textwidth]{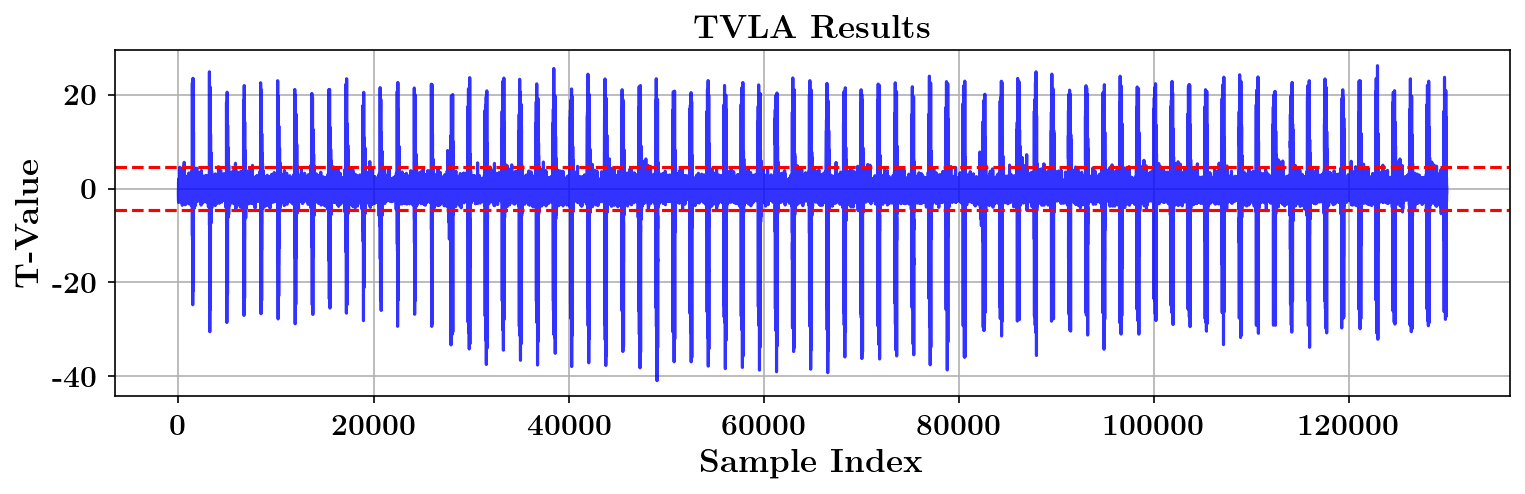}
      \caption{TVLA leakage analysis: The red line represents the t-test threshold at $\pm 4.5$.}
        \label{fig:class1}
    \end{subfigure}\\[0.3cm]
    \begin{subfigure}[b]{0.8\textwidth}
        \centering
        \includegraphics[width=\textwidth]{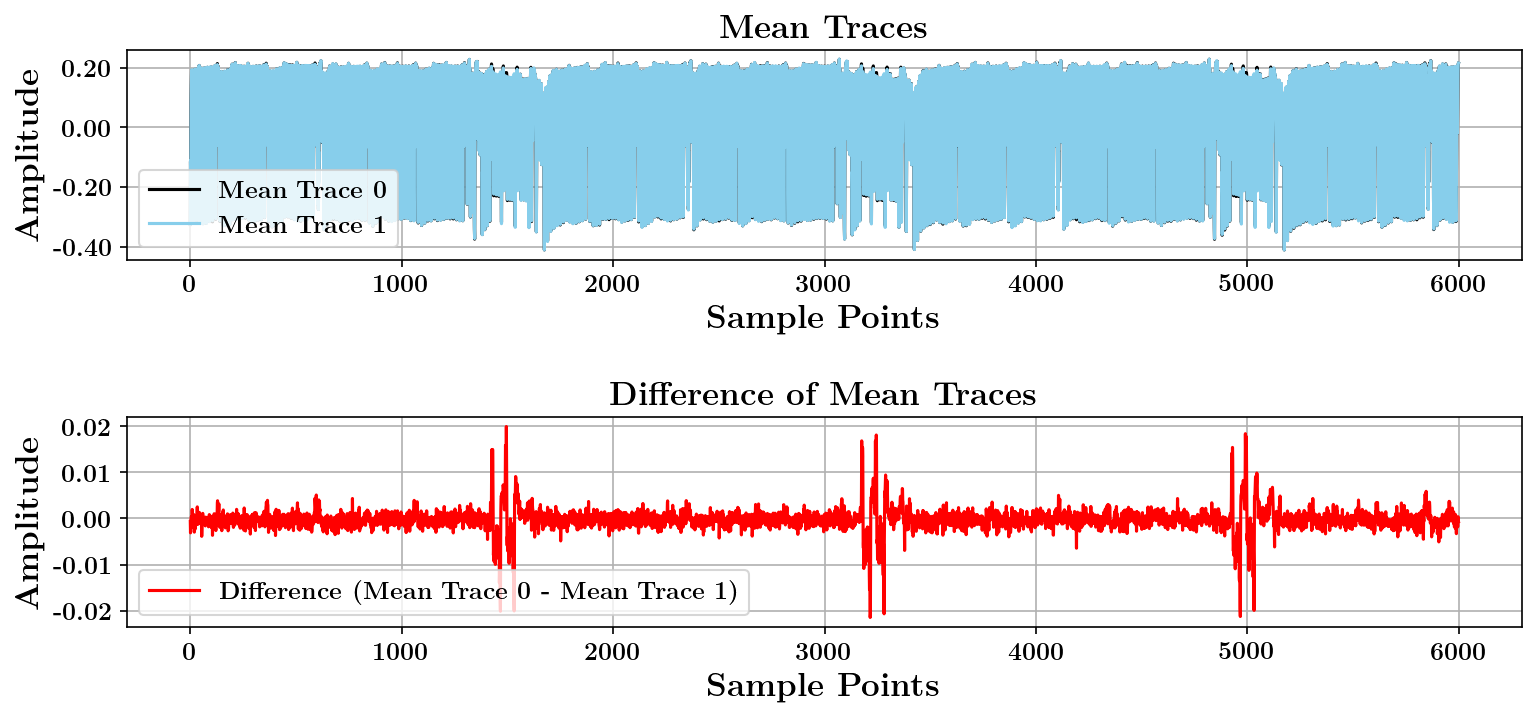}
        \caption{Mean traces and the difference of mean traces.}
        \label{fig:class2}
    \end{subfigure}
\caption{Leakage analysis for the encoding function during re-encryption in \protect\KEMD. Each figure was plotted using data from \numprint{200} power traces per group. Only partial traces are shown, starting after \numprint{300000} clock cycles. Figure~\protect\ref{fig:class1} includes a total of \numprint{130000} consecutive data points, while Figure~\protect\ref{fig:class2} includes \numprint{6000} consecutive data points.}
    \label{fig:traces}
\end{figure}

\paragraph{Leakage Detection}

Previous studies have identified several sources of power leakage that could be utilized to construct the required PC oracle for HQC. Notable examples include power side-channel leakages during hash function computation in the re-encryption process, as reported in~\cite{TCHES:UXTITH22}, and from the \RS decoder, as detailed in~\cite{DBLP:conf/pqcrypto/SchambergerHRWS22}. Our research focuses on power traces from the \RM\ \RS encoding procedures during the re-encryption of HQC’s decapsulation \KEMD phase. This extensive computation phase, primarily dependent on the input $m'$, provides significant leakages, making it ideal for constructing a PC oracle to ascertain decryption outcomes.

In our experiments, the complete trace of this \RM\ \RS encoding process during HQC re-encryption spans over \numprint{800000} clock cycles. We analyze \numprint{130000} data points starting from the \numprint{300000}th timestamp for each trace. To pinpoint leakage and identify the most significant leakage points, we conduct the Test Vector Leakage Assessment (TVLA) on 200 traces of decryption successes (decrypting to all zeros) and 200 traces of decryption failures. The TVLA results are illustrated in Figure~\protect\ref{fig:class1}, with a detailed analysis of the initial \numprint{6000} data points presented in Figure~\protect\ref{fig:class2} to highlight the trace differences. We select sample positions exhibiting t-values greater than 16 in the TVLA test as the strongest leakage points, resulting in \numprint{1280} points per trace.

\paragraph{Experimental Results}

\begin{table}[t]
    \centering
    \caption{CNN hyperparameters (adapted from~\cite{TCHES:UXTITH22})}
\small
    \scalebox{0.85}{
    \begin{tabular}{cccccc}
      \toprule
         &Operator&Activation function&Batch normalization&Pooling&Stride\\
             \midrule
         Conv1& conv1d (3)& SELU& Yes& Avg (2)&2\\
         Conv2& conv1d (3)& SELU& Yes& Avg (2)&2\\
         Conv3& conv1d (3)& SELU& Yes& Avg (2)&2\\
         Conv4& conv1d (3)& SELU& Yes& Avg (2)&2\\
         Conv5& conv1d (3)& SELU& Yes& Avg (2)&2\\
         Conv6& conv1d (3)& SELU& Yes& Avg (2)&2\\
         FLT& flatten& -& -& -&-\\
          FC1& dense& SELU& No&  No&-\\
          FC2& dense& SELU&  No&  No&-\\
           FC3& dense& Sigmoid &  No&  No&-\\
       
            \bottomrule
    \end{tabular}
    }

    \label{tab:hyper}
\end{table}

We employ a convolutional neural network (CNN) architecture, as outlined in Table~\ref{tab:hyper}, to train a model capable of classifying whether decryption is successful. This model is adapted from~\cite{TCHES:UXTITH22}, with adjustments made to the input size accordingly. We capture traces and retain only the \numprint{1280} data points corresponding to positions that exhibit a t-value greater than 16 in our TVLA test, resulting in an input vector of dimension \numprint{1280}.

During the training phase, we collect traces from 80 runs of the attack, each using a different newly generated key pair and seven predefined patterns (i.e., $t=7$). Each key pair produces 644 traces, calculated as $46\times 7\times 2$, so we collect a total of \numprint{51520} traces. We separate the dataset into training, validation, and testing sets consisting of \numprint{30418}, \numprint{10138}, and \numprint{10138} traces, respectively. To ensure balanced samples between the groups of decryption success and decryption failure, some traces are excluded based on the decrypted messages as ground truth.
We train the model over 100 epochs. The trained model achieves an accuracy of 1 on the test set, indicating no classification errors on the \numprint{10138} test traces. The model training is performed on a MacBook Pro laptop with a 10-core M1 Pro CPU, a 14-core built-in GPU, and 32GB of RAM.

We then apply the trained model to \numprint{77280} traces collected from 120 newly generated random key pairs to ascertain decryption outcomes. Using the classification results, we employ the offline template designed for a perfect oracle (\(\rho=1\)) to compute the conditional probabilities of each secret entry being 1, similar to our earlier simulations. Subsequently, the bits are ranked based on these probabilities. The average number of errors and the success rates of the attack are presented in Table \ref{tab:realw}, alongside the simulation results. The results from the real-world validation and the simulations align perfectly.

\begin{table}[t!]
\centering
\small
    \setlength\tabcolsep{0.25cm} 
\caption{Comparison of simulation and real-world validation with seven PC oracle calls per length-\(n_2\) block (\(t=7\)) for \protect\hqcA. The validation includes 120 real-world tests, each conducted with a different newly generated key pair.}
\label{tab:realw}
\begin{tabular}{@{}lccc@{}}
\hline
                      & \multirow{2}{*}{\begin{tabular}[c]{@{}c@{}}Mean Number \\ of Errors\end{tabular}} & \multicolumn{2}{c}{Success Rate (\Psucc)} \\ \cline{3-4} 
                      &                                                                                   & $w_0=2$            & $w_0=3$           \\ \hline
                      
Simulation ($\rho=1$)    & 2.25                                                                              & 85.45\%         & 92.60\%        \\
Real-world validation & 2.35                                                                              & 84.17\%         & 92.50\%        \\ \hline
\end{tabular}
\end{table}
\section{Discussion~\label{sec:dis}}
In this section, we discuss the gains of our attack methodology compared to the state-of-the-art \ZT~method, and its potential use in constructing multi-value or parallel PC oracle based side-channel attacks on HQC.

\subsection{More on Comparison with the \ZT~Method
~\label{sec:MoreZT}}
In Section~\ref{sec:sim}, we demonstrated that our new method requires significantly fewer oracle calls compared to the previous best, the \ZT~method~\cite{USS:SchroderGG24}. Additionally, our method is more robust to the inaccuracy in the constructed PC oracle.

Another notable advantage of our approach over the \ZT~method is its ability to address a limitation discussed in Section 6.3 of~\cite{USS:SchroderGG24}. The effectiveness of the \ZT~method depends on identifying long sequences of consecutive zero entries between two ones in the secret polynomials $x$ and $y$. Even with an increased number of side-channel measurements, the \ZT~method is effective for only a small proportion (approximately 15\%) of keys for \hqcB, where the likelihood of finding an all-zero block of length $n_2$ within the secret vectors $x$ and $y$ is low.

\begin{figure}[t] 
\centering
\includegraphics[width=0.5\textwidth]{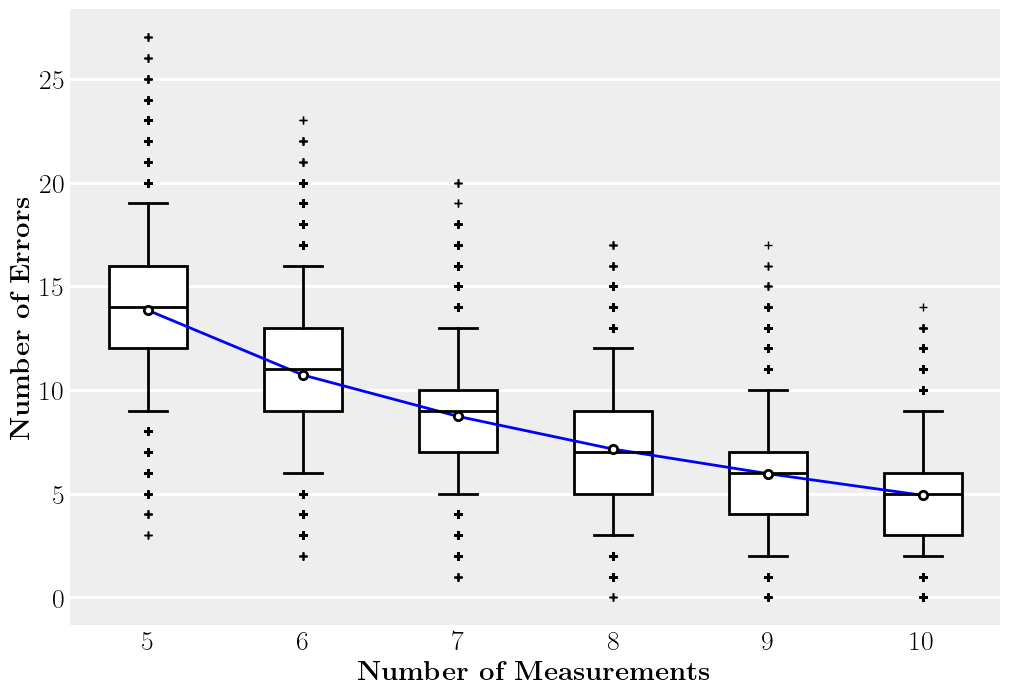}
\caption{The number of errors in the first $n$ positions of the reordered $\vect z$ vector decreases as the number of side-channel measurements per length-$n_2$ block increases for \hqcB~with an oracle accuracy of $\rho = 1$. The plot is based on testing \numprint{10000} random keys for each scenario. The whiskers represent the 5th and 95th percentiles. The blue line connects the mean number of errors.}
\label{fig:hqcB}
\end{figure}

As shown in Figure~\ref{fig:hqcB}, our method effectively reduces the number of decision errors in the first $n$ positions of the reordered $\vect z$ vector as the number of side-channel measurements increases for \hqcB. The x-axis represents the number of measurements per \RM~block ($t$). Detailed information on sample complexity, post-processing complexity, and the probability of full key recovery for \hqcB~with an oracle accuracy of $\rho = 1$ is provided in Section~\ref{sec:sim}.

\subsection{Extension to Multi-value or Parallel PC oracle~\label{sec:mul}}
Rajendran et al.~\cite{TCHES:RRDBC23} and Tanaka et al.~\cite{TCHES:TUXITH23} were the first to propose side-channel attacks based on multi-value or parallel PC oracles. In~\cite{TCHES:RRDBC23}, the oracle is referred to as a parallel PC oracle, while in~\cite{TCHES:TUXITH23}, it is termed a multi-value oracle. The concept behind these oracles is that, in specific scenarios such as unprotected software implementations on low-end platforms, power or EM leakages are substantial enough to construct a multi-value classifier. This allows a multi-value or parallel PC oracle to leak more than one bit of information, thereby significantly reducing the number of required traces. While these attacks have limited applications, they are more powerful than binary PC oracle based attacks when a multi-value or parallel PC oracle can be built. As noted by Tanaka et al. in~\cite{TCHES:TUXITH23}, constructing such attacks for HQC remains an open problem.

The new OT-PCA method can be extended to address this open problem by creating different offline templates for more than two decrypted messages. 
Preliminary investigations have demonstrated gains compared to binary PC oracle based side-channel attacks. However, extensive experiments are necessary to evaluate and optimize these concrete performance gains. We leave this research problem for future work because multi-value or parallel PC oracles are important oracles that deserve further and separate investigation. 

\section{Concluding Remarks}
\label{sec:conclusion}

We have introduced OT-PCA, a novel key-recovery PC oracle based side-channel attack against HQC, utilizing templates constructed from offline access to the publicly available decoding function of \RM~codes. These templates provide valuable soft information for the post-processing stage. 
Our method requires minimal post-processing costs, comparable to Gaussian elimination. Simulation results show that OT-PCA significantly reduces the required number of PC oracle calls compared to state-of-the-art attacks. Furthermore, this new attack is more robust to inaccuracy in the PC oracle and performs significantly better than the method in \cite{USS:SchroderGG24} on \hqcB.
The effectiveness of this approach has also been validated through real-world tests using power traces on a ChipWhisperer Husky platform equipped with an ARM Cortex-M4 CPU, confirming the simulation results and showcasing the practical relevance of our method.

\myred{Countermeasures like constant-time implementation, masking, or shuffling can make constructing accurate PC oracles difficult or infeasible, thus hardening the attack.}
Future work includes a comprehensive evaluation of multi-value or parallel PC oracle based side-channel attacks using the OT-PCA method on HQC to understand its efficiency and potential limitations. Additionally, evaluating software PC oracle based side-channel attacks, such as GoFetch~\cite{gofetch}, against HQC will be valuable. Last, extending the OT-PCA method to Kyber and other lattice-based cryptographic schemes will provide insights into the implementation security of various lattice-based proposals.

\section*{Acknowledgement}
This work was supported by the Swedish Research Council (grant numbers 2021-04602 and 2023-03654), the Swedish Civil Contingencies Agency (grant number 2020-11632), the Swedish Foundation for Strategic Research (grant number RIT17-0005), the Crafoord Foundation, and the Wallenberg AI, Autonomous Systems and Software Program (WASP), funded by the Knut and Alice Wallenberg Foundation. The computations and simulations were partly enabled by resources provided by LUNARC.

\bibliographystyle{alpha}
\bibliography{abbrev3,crypto,other}

\newcommand{\etalchar}[1]{$^{#1}$}
\begin{thebibliography}{WTBB{\etalchar{+}}19}

\bibitem[AAB{\etalchar{+}}22]{NISTPQC-R4:HQC22}
Carlos {Aguilar-Melchor}, Nicolas Aragon, Slim Bettaieb, Lo{\"i}c Bidoux, Olivier Blazy, Jean-Christophe Deneuville, Philippe Gaborit, Edoardo Persichetti, Gilles Z{\'e}mor, Jurjen Bos, Arnaud Dion, Jerome Lacan, Jean-Marc Robert, and Pascal Veron.
\newblock {HQC}.
\newblock Technical report, {N}ational {I}nstitute of {S}tandards and {T}echnology, 2022.
\newblock available at \url{https://csrc.nist.gov/Projects/post-quantum-cryptography/round-4-submissions}.

\bibitem[ABB{\etalchar{+}}22]{NISTPQC-R4:BIKE22}
Nicolas Aragon, Paulo Barreto, Slim Bettaieb, Loic Bidoux, Olivier Blazy, Jean-Christophe Deneuville, Phillipe Gaborit, Shay Gueron, Tim Guneysu, Carlos {Aguilar-Melchor}, Rafael Misoczki, Edoardo Persichetti, Nicolas Sendrier, Jean-Pierre Tillich, Gilles Z{\'e}mor, Valentin Vasseur, Santosh Ghosh, and Jan Richter-Brokmann.
\newblock {BIKE}.
\newblock Technical report, {N}ational {I}nstitute of {S}tandards and {T}echnology, 2022.
\newblock available at \url{https://csrc.nist.gov/Projects/post-quantum-cryptography/round-4-submissions}.

\bibitem[ABC{\etalchar{+}}22]{NISTPQC-R4:ClassicMcEliece22}
Martin~R. Albrecht, Daniel~J. Bernstein, Tung Chou, Carlos Cid, Jan Gilcher, Tanja Lange, Varun Maram, Ingo von Maurich, Rafael Misoczki, Ruben Niederhagen, Kenneth~G. Paterson, Edoardo Persichetti, Christiane Peters, Peter Schwabe, Nicolas Sendrier, Jakub Szefer, Cen~Jung Tjhai, Martin Tomlinson, and Wen Wang.
\newblock {Classic McEliece}.
\newblock Technical report, {N}ational {I}nstitute of {S}tandards and {T}echnology, 2022.
\newblock available at \url{https://csrc.nist.gov/projects/post-quantum-cryptography/round-4-submissions}.

\bibitem[ACD{\etalchar{+}}22]{915326}
Gorjan Alagic, David Cooper, Quynh Dang, Thinh Dang, John~M. Kelsey, Jacob Lichtinger, Yi-Kai Liu, Carl~A. Miller, Dustin Moody, Rene Peralta, Ray Perlner, Angela Robinson, Daniel Smith-Tone, and Daniel Apon.
\newblock Status report on the third round of the nist post-quantum cryptography standardization process, 2022-07-05 04:07:00 2022.

\bibitem[BDH{\etalchar{+}}19]{EC:BDHTV19}
Ciprian B{\u a}etu, F.~Bet{\"u}l Durak, Lo{\"i}s {Huguenin-Dumittan}, Abdullah Talayhan, and Serge Vaudenay.
\newblock Misuse attacks on post-quantum cryptosystems.
\newblock In Yuval Ishai and Vincent Rijmen, editors, {\em EUROCRYPT~2019, Part~II}, volume 11477 of {\em {LNCS}}, pages 747--776. Springer, Cham, May 2019.

\bibitem[CWS{\etalchar{+}}24]{gofetch}
Boru Chen, Yingchen Wang, Pradyumna Shome, Christopher~W. Fletcher, David Kohlbrenner, Riccardo Paccagnella, and Daniel Genkin.
\newblock Gofetch: Breaking constant-time cryptographic implementations using data memory-dependent prefetchers.
\newblock In {\em USENIX Security}, 2024.

\bibitem[FO99]{C:FujOka99}
Eiichiro Fujisaki and Tatsuaki Okamoto.
\newblock Secure integration of asymmetric and symmetric encryption schemes.
\newblock In Michael~J. Wiener, editor, {\em CRYPTO'99}, volume 1666 of {\em {LNCS}}, pages 537--554. Springer, Berlin, Heidelberg, August 1999.

\bibitem[FS09]{AC:FinSen09}
Matthieu Finiasz and Nicolas Sendrier.
\newblock Security bounds for the design of code-based cryptosystems.
\newblock In Mitsuru Matsui, editor, {\em ASIACRYPT~2009}, volume 5912 of {\em {LNCS}}, pages 88--105. Springer, Berlin, Heidelberg, December 2009.

\bibitem[GHJ{\etalchar{+}}22]{TCHES:GHJLNS22}
Qian Guo, Clemens Hlauschek, Thomas Johansson, Norman Lahr, Alexander Nilsson, and Robin~Leander Schr{\"o}der.
\newblock Don't reject this: Key-recovery timing attacks due to rejection-sampling in {HQC} and {BIKE}.
\newblock {\em {IACR} {TCHES}}, 2022(3):223--263, 2022.

\bibitem[GJ20]{AC:GuoJoh20}
Qian Guo and Thomas Johansson.
\newblock A new decryption failure attack against {HQC}.
\newblock In Shiho Moriai and Huaxiong Wang, editors, {\em ASIACRYPT~2020, Part~I}, volume 12491 of {\em {LNCS}}, pages 353--382. Springer, Cham, December 2020.

\bibitem[GJMW19]{guo2019some}
Qian Guo, Thomas Johansson, Erik M{\aa}rtensson, and Paul~Stankovski Wagner.
\newblock Some cryptanalytic and coding-theoretic applications of a soft stern algorithm.
\newblock {\em Advances in Mathematics of Communications}, 13(4), 2019.

\bibitem[GJN20]{C:GuoJohNil20}
Qian Guo, Thomas Johansson, and Alexander Nilsson.
\newblock A key-recovery timing attack on post-quantum primitives using the {Fujisaki}-{Okamoto} transformation and its application on {FrodoKEM}.
\newblock In Daniele Micciancio and Thomas Ristenpart, editors, {\em CRYPTO~2020, Part~II}, volume 12171 of {\em {LNCS}}, pages 359--386. Springer, Cham, August 2020.

\bibitem[GLG22]{DBLP:conf/pqcrypto/GoyLG22}
Guillaume Goy, Antoine Loiseau, and Philippe Gaborit.
\newblock A new key recovery side-channel attack on {HQC} with chosen ciphertext.
\newblock In Jung~Hee Cheon and Thomas Johansson, editors, {\em Post-Quantum Cryptography - 13th International Workshop, PQCrypto 2022, Virtual Event, September 28-30, 2022, Proceedings}, volume 13512 of {\em Lecture Notes in Computer Science}, pages 353--371. Springer, 2022.

\bibitem[GMGL24]{DBLP:journals/tches/GoyMGL24}
Guillaume Goy, Julien Maillard, Philippe Gaborit, and Antoine Loiseau.
\newblock Single trace {HQC} shared key recovery with {SASCA}.
\newblock {\em {IACR} Trans. Cryptogr. Hardw. Embed. Syst.}, 2024(2):64--87, 2024.

\bibitem[GNNJ23]{AC:GNNJ23}
Qian Guo, Denis Nabokov, Alexander Nilsson, and Thomas Johansson.
\newblock {SCA}-{LDPC}: {A} code-based framework for key-recovery side-channel attacks on post-quantum encryption schemes.
\newblock In Jian Guo and Ron Steinfeld, editors, {\em ASIACRYPT~2023, Part~IV}, volume 14441 of {\em {LNCS}}, pages 203--236. Springer, Singapore, December 2023.

\bibitem[HHK17]{TCC:HofHovKil17}
Dennis Hofheinz, Kathrin H{\"o}velmanns, and Eike Kiltz.
\newblock A modular analysis of the {Fujisaki}-{Okamoto} transformation.
\newblock In Yael Kalai and Leonid Reyzin, editors, {\em TCC~2017, Part~I}, volume 10677 of {\em {LNCS}}, pages 341--371. Springer, Cham, November 2017.

\bibitem[HPP21]{DBLP:conf/indocrypt/HermelinkPP21}
Julius Hermelink, Peter Pessl, and Thomas P{\"{o}}ppelmann.
\newblock Fault-enabled chosen-ciphertext attacks on kyber.
\newblock In Avishek Adhikari, Ralf K{\"{u}}sters, and Bart Preneel, editors, {\em Progress in Cryptology - {INDOCRYPT} 2021 - 22nd International Conference on Cryptology in India, Jaipur, India, December 12-15, 2021, Proceedings}, volume 13143 of {\em Lecture Notes in Computer Science}, pages 311--334. Springer, 2021.

\bibitem[HSC{\etalchar{+}}23]{TCHES:HSCGJ23}
Senyang Huang, Rui~Qi Sim, Chitchanok Chuengsatiansup, Qian Guo, and Thomas Johansson.
\newblock Cache-timing attack against {HQC}.
\newblock {\em {IACR} {TCHES}}, 2023(3):136--163, 2023.

\bibitem[HV20]{ACNS:HugVau20}
Lo{\"i}s {Huguenin-Dumittan} and Serge Vaudenay.
\newblock Classical misuse attacks on {NIST} round 2 {PQC} - the power of rank-based schemes.
\newblock In Mauro Conti, Jianying Zhou, Emiliano Casalicchio, and Angelo Spognardi, editors, {\em ACNS 20International Conference on Applied Cryptography and Network Security, Part~I}, volume 12146 of {\em {LNCS}}, pages 208--227. Springer, Cham, October 2020.

\bibitem[JAC{\etalchar{+}}22]{NISTPQC-R4:SIKE22}
David Jao, Reza Azarderakhsh, Matthew Campagna, Craig Costello, Luca {De Feo}, Basil Hess, Amir Jalali, Brian Koziel, Brian {LaMacchia}, Patrick Longa, Michael Naehrig, Joost Renes, Vladimir Soukharev, David Urbanik, Geovandro Pereira, Koray Karabina, and Aaron Hutchinson.
\newblock {SIKE}.
\newblock Technical report, {N}ational {I}nstitute of {S}tandards and {T}echnology, 2022.
\newblock available at \url{https://csrc.nist.gov/Projects/post-quantum-cryptography/round-4-submissions}.

\bibitem[Meu13]{meurer2013coding}
Alexander Meurer.
\newblock {\em A coding-theoretic approach to cryptanalysis}.
\newblock PhD thesis, Verlag nicht ermittelbar, 2013.

\bibitem[MKB{\etalchar{+}}24]{ACNS:MKBKV24}
Puja Mondal, Suparna Kundu, Sarani Bhattacharya, Angshuman Karmakar, and Ingrid Verbauwhede.
\newblock A practical key-recovery attack on {LWE}-based key-encapsulation mechanism schemes using rowhammer.
\newblock In Christina P{\"o}pper and Lejla Batina, editors, {\em ACNS 24International Conference on Applied Cryptography and Network Security, Part~III}, volume 14585 of {\em {LNCS}}, pages 271--300. Springer, Cham, March 2024.

\bibitem[PRJB23]{EPRINT:PRJB23}
Thales Paiva, Prasanna Ravi, Dirmanto Jap, and Shivam Bhasin.
\newblock Et tu, brute? {SCA} assisted {CCA} using valid ciphertexts - {A} case study on {HQC} {KEM}.
\newblock Cryptology {ePrint} Archive, Report 2023/1626, 2023.

\bibitem[PT19]{SAC:PaiTer19}
Thales~Bandiera Paiva and Routo Terada.
\newblock A timing attack on the {HQC} encryption scheme.
\newblock In Kenneth~G. Paterson and Douglas Stebila, editors, {\em SAC 2019}, volume 11959 of {\em {LNCS}}, pages 551--573. Springer, Cham, August 2019.

\bibitem[RRCB20]{TCHES:RRCB20}
Prasanna Ravi, Sujoy~Sinha Roy, Anupam Chattopadhyay, and Shivam Bhasin.
\newblock Generic side-channel attacks on {CCA}-secure lattice-based {PKE} and {KEMs}.
\newblock {\em {IACR} {TCHES}}, 2020(3):307--335, 2020.

\bibitem[RRD{\etalchar{+}}23]{TCHES:RRDBC23}
Gokulnath Rajendran, Prasanna Ravi, Jan-Pieter D'Anvers, Shivam Bhasin, and Anupam Chattopadhyay.
\newblock Pushing the limits of generic side-channel attacks on {LWE}-based {KEMs} - parallel {PC} oracle attacks on {Kyber} {KEM} and beyond.
\newblock {\em {IACR} {TCHES}}, 2023(2):418--446, 2023.

\bibitem[SAB{\etalchar{+}}22]{NISTPQC:CRYSTALS-KYBER22}
Peter Schwabe, Roberto Avanzi, Joppe Bos, L{\'e}o Ducas, Eike Kiltz, Tancr{\`e}de Lepoint, Vadim Lyubashevsky, John~M. Schanck, Gregor Seiler, Damien Stehl{\'e}, and Jintai Ding.
\newblock {CRYSTALS-KYBER}.
\newblock Technical report, {N}ational {I}nstitute of {S}tandards and {T}echnology, 2022.
\newblock available at \url{https://csrc.nist.gov/Projects/post-quantum-cryptography/selected-algorithms-2022}.

\bibitem[SCZ{\etalchar{+}}23]{TCHES:SCZGJ23}
Muyan Shen, Chi Cheng, Xiaohan Zhang, Qian Guo, and Tao Jiang.
\newblock Find the bad apples: An efficient method for perfect key recovery under imperfect {SCA} oracles - {A} case study of {Kyber}.
\newblock {\em {IACR} {TCHES}}, 2023(1):89--112, 2023.

\bibitem[SGG24]{USS:SchroderGG24}
Robin~Leander Schr{\"{o}}der, Stefan Gast, and Qian Guo.
\newblock Divide and surrender: Exploiting variable division instruction timing in {HQC} key recovery attacks.
\newblock In {\em {USENIX} Security}. {USENIX} Association, 2024.

\bibitem[SHR{\etalchar{+}}22]{DBLP:conf/pqcrypto/SchambergerHRWS22}
Thomas Schamberger, Lukas Holzbaur, Julian Renner, Antonia Wachter{-}Zeh, and Georg Sigl.
\newblock {A Power Side-Channel Attack on the Reed-Muller Reed-Solomon Version of the {HQC} Cryptosystem}.
\newblock In Jung~Hee Cheon and Thomas Johansson, editors, {\em Post-Quantum Cryptography - 13th International Workshop, PQCrypto 2022, Virtual Event, September 28-30, 2022, Proceedings}, volume 13512 of {\em Lecture Notes in Computer Science}, pages 327--352. Springer, 2022.

\bibitem[TUX{\etalchar{+}}23]{TCHES:TUXITH23}
Yutaro Tanaka, Rei Ueno, Keita Xagawa, Akira Ito, Junko Takahashi, and Naofumi Homma.
\newblock Multiple-valued plaintext-checking side-channel attacks on post-quantum {KEMs}.
\newblock {\em {IACR} {TCHES}}, 2023(3):473--503, 2023.

\bibitem[UXT{\etalchar{+}}22]{TCHES:UXTITH22}
Rei Ueno, Keita Xagawa, Yutaro Tanaka, Akira Ito, Junko Takahashi, and Naofumi Homma.
\newblock Curse of re-encryption: {A} generic power/{EM} analysis on post-quantum {KEMs}.
\newblock {\em {IACR} {TCHES}}, 2022(1):296--322, 2022.

\bibitem[WPH{\etalchar{+}}22]{USENIX:WPHSFK22}
Yingchen Wang, Riccardo Paccagnella, Elizabeth~Tang He, Hovav Shacham, Christopher~W. Fletcher, and David Kohlbrenner.
\newblock Hertzbleed: Turning power side-channel attacks into remote timing attacks on {x86}.
\newblock In Kevin R.~B. Butler and Kurt Thomas, editors, {\em USENIX Security 2022}, pages 679--697. {USENIX} Association, August 2022.

\bibitem[WTBB{\etalchar{+}}19]{EPRINT:WBBGM19}
Guillaume Wafo-Tapa, Slim Bettaieb, Loic Bidoux, Philippe Gaborit, and Etienne Marcatel.
\newblock A practicable timing attack against {HQC} and its countermeasure.
\newblock Cryptology ePrint Archive, Report 2019/909, 2019.

\bibitem[XIU{\etalchar{+}}21]{AC:XIUTH21}
Keita Xagawa, Akira Ito, Rei Ueno, Junko Takahashi, and Naofumi Homma.
\newblock Fault-injection attacks against {NIST}'s post-quantum cryptography round 3 {KEM} candidates.
\newblock In Mehdi Tibouchi and Huaxiong Wang, editors, {\em ASIACRYPT~2021, Part~II}, volume 13091 of {\em {LNCS}}, pages 33--61. Springer, Cham, December 2021.

\end{thebibliography}

\end{document}